\documentclass[%
 reprint,
 nofootinbib,
 amsmath,amssymb,
 aps,
 prd,
 floatfix
]{revtex4-2}

\usepackage{graphicx}% Include figure files
\usepackage[caption=false]{subfig}
\usepackage{tabularx}
\usepackage{float}
\usepackage[separate-uncertainty=true,multi-part-units=single]{siunitx}
\usepackage{hyperref}% add hypertext capabilities
\usepackage{cleveref}
\usepackage{amsmath}
\usepackage{caption}
\usepackage{subcaption}
\usepackage{braket}
\usepackage{tensor}
\usepackage{booktabs}
\usepackage[nolist,nohyperlinks]{acronym}
\usepackage{enumerate}
\usepackage{orcidlink}
\DeclareFontFamily{U}{dutchcal}{\skewchar\font=45 }
\DeclareFontShape{U}{dutchcal}{m}{n}{<-> s*[1.0] dutchcal-r}{}
\DeclareFontShape{U}{dutchcal}{b}{n}{<-> s*[1.0] dutchcal-b}{}
\DeclareMathAlphabet{\mathlcal}{U}{dutchcal}{m}{n}
\SetMathAlphabet{\mathlcal}{bold}{U}{dutchcal}{b}{n}
\DeclareSIUnit \parsec {pc}
\DeclareSIUnit \year{yr}
\DeclareSIUnit\pc{pc}

\newcommand{\crefrangeconjunction}{--}

\crefrangelabelformat{equation}{(#3#1#4 \crefrangeconjunction ~#5#2#6)}

\begin{document}

\crefname{section}{Sec.}{Sec.}
\crefname{equation}{Eq.}{Eqs.}
\crefname{figure}{Fig.}{Fig.}
\crefname{table}{Table}{Table}
\crefname{appendix}{Appx.}{Appx.}

\title{Bianchi Type I Model Cannot Explain the Observed CMB Angular Acoustic Scale Directional Variation}

\author{Boris Hoi-Lun Ng~\orcidlink{0009-0000-4811-5294}}
\email{boris.hl.ng@link.cuhk.edu.hk}
\affiliation{%
Department of Physics, The Chinese University of Hong Kong, Shatin, Hong Kong
}%
\author{Ming-Chung Chu~\orcidlink{0000-0002-1971-0403}}
\email{mcchu@phy.cuhk.edu.hk}
\affiliation{%
Department of Physics, The Chinese University of Hong Kong, Shatin, Hong Kong
}%
\date{\today}
\begin{abstract}
Anisotropic cosmological models have been gaining attention due to various observational hints of large-scale anisotropies. One of the most surprising evidences for the latter is the discovery of a dipole-like directional variation in cosmological parameters extracted from the Cosmic Microwave Background (CMB) data. In this work, we show that the directional variation of the CMB angular acoustic angle calculated with the fully asymmetric Bianchi Type I metric, a simple extension of the standard Friedmann–Lema\^itre–Robertson–Walker metric, cannot account for the observed dipole-like anisotropy.
\end{abstract}
\maketitle

\begin{acronym}
    \acro{CP}{Cosmological Principle}
    \acro{FLRW}{Friedmann–Lema\^itre–Robertson–Walker}
    \acro{CMB}{Cosmic Microwave Background}
    \acro{BAO}{Baryon Acoustic Oscillation}
    \acro{MCMC}{Markov Chain Monte Carlo}
    \acro{CvB}[C$\nu$B]{Cosmic Neutrino Background}
\end{acronym}
\section{\label{Intro}Introduction}
Recent cosmological observations have challenged the foundations of modern cosmological theories, the \ac{CP} of homogeneity and isotropy on large scales. In particular, the \ac{CMB} anisotropies were found to exhibit a North-South asymmetry \cite{Hansen_2004, Eriksen_2004, Bernui_2006, Bernui_2008} and alignments of some multipoles \cite{Bielewicz_2005, deOliveira-Costa_2004, Tegmark_2003, Wiaux_2006, Naselsky_2012, Zhao_2014, Cheng_2016, Aluri_2017}, with some authors even claiming a preferred axis \cite{Bunn_2000, Hansen_2004, Eriksen_2004, Bernui_2007, Hansen_2009} in \ac{CMB}. Interestingly, these anomalies roughly align with anisotropic features on local scales \cite{Kashlinsky_2008, Watkins_2009, Hutsemekers_2005}, such as that for the Hubble constant $H_0$ in Type Ia Supernovae measurements \cite{Javanmardi_2015, Krishnan_2022, Luongo_2022, Hu_2023}. Refs. \cite{Antoniou_2010, Zhao_2016, Perivolaropoulos_2022} provide comprehensive reviews of these anomalies.

A more direct piece of evidence for an anisotropic universe is the discovery of directionally dependent cosmological parameters extracted from \ac{CMB} data. Using the Wilkinson Microwave Anisotropy Probe data, Ref. \cite{Axelsson_2013} demonstrated a directional variation in $n_s$, $A_s$, and $\Omega_{b0}$. Ref. \cite{Yeung_2022} performed similar analyses with the more precise Planck data and discovered anisotropy in all six cosmological parameters $\set{\Omega_{b} h^2, \Omega_{c} h^2, n_s, 100\theta_\mathrm{MC}, \tau, \ln{(10^{10}A_s)}}$ in addition to $H_0$. This dipole-like variation in the cosmological parameters remains one of the strongest evidences for an anisotropic universe. 

Given the apparent violation of \ac{CP} and the fact that the standard \ac{FLRW} metric inherently assumes \ac{CP}, it becomes necessary to explore alternative metrics. A commonly considered alternative is the Bianchi Type I model
\begin{gather}
    ds^2 = -dt^2 + a^2(t)dx^2 + b^2(t)dy^2 + c^2(t)dz^2, \label{Bianchi metric}
\end{gather}
a natural anisotropic extension of the \ac{FLRW} metric that allows for large-scale anisotropy while maintaining homogeneity \cite{Akarsu_2019,Shekh_2020,Sarmah_2022,Koussour_2023,Hertzberg_Loeb_2024,Ng_2025}. Here, $a(t)$, $b(t)$, and $c(t)$ are the scale factors in the $x$, $y$, and $z$ directions, respectively, replacing the \ac{FLRW} scale factor. Trivially, one recovers the \ac{FLRW} metric when $a=b=c$. In our previous work \cite{Ng_2025}, we showed that a locally rotationally symmetric Bianchi Type I model can produce a directional variation in the angular acoustic scale of \ac{CMB} $\theta_*$. In this work, we extend our study by considering a fully asymmetric Bianchi Type I model. Moreover, we adhere to the methods used in Ref. \cite{Yeung_2022} for a fair comparison with observations.

In \cref{Model}, we first introduce the Bianchi Type I model and how the angular acoustic scale $\theta_*$ is calculated. Next, we discuss our methodology in \cref{Method}. The results and comparison with observations are presented in \cref{Results}. Finally, we conclude in \cref{Conclusion}.

\section{The Anisotropic universe \label{Model}}
\subsection{Bianchi Type I Model \label{GR}}
In the comoving frame, the line element of the Bianchi Type I model is defined by \cref{Bianchi metric} \cite{Akarsu_2019,Yadav_2023,Hertzberg_Loeb_2024,Sarmah_2022,Koivisto_2008,Ng_2025}. Note that geometric units such that $c=G=1$ are assumed throughout this work. Similar to the case for the \ac{FLRW} metric, one can define the corresponding Hubble parameters as
\begin{align}
    H_x&=\frac{\dot{a}}{a}, &
    H_y&=\frac{\dot{b}}{b}, &
    H_z&=\frac{\dot{c}}{c},
\end{align}
where the dot represents a derivative with respect to time.

Meanwhile, by treating radiation ($r$, which includes massless neutrinos), matter ($m$), and dark energy (in the form of the cosmological constant $\Lambda$) as perfect fluids with equations of state $p_X=w_X\rho_X$, $X=r,\ m, \text{ or } \Lambda$, the energy-momentum tensor $T\indices{^\mu_{\nu}}$ is given by
\begin{align}
    T\indices{^\mu_{\nu}}=\text{diag}(-\rho, p, p, p).
\end{align}
Note that an isotropic pressure is assumed \cite{Akarsu_2019,Sarmah_2022,Yadav_2023,Hertzberg_Loeb_2024,Ng_2025}. Here, $\rho$ and $p$ include contributions from all three fluids, $\rho=\sum \rho_X$ and $p=\sum p_X=\sum w_X \rho_X$, where $(w_r, w_m, w_\Lambda)$ equal to $(1/3, 0, -1)$. The evolution of energy densities is governed by the local conservation law
\begin{gather}
    \nabla_\mu T\indices{^{\mu\nu}}=0,
\intertext{or}
    \dot{\rho}_X+(1+w_X)\rho_X(H_x+H_y+H_z)=0 \label{d_rho},\\
    \rho_X = \rho_{X0}\left(\frac{abc}{a_0 b_0 c_0}\right)^{-(1+w_X)}. \label{rho_a}
\end{gather}

To solve the evolution of the scale factors, one considers the Einstein Field Equation
\begin{align}
    G\indices{_{\mu\nu}}=8\pi T\indices{_{\mu\nu}},
\end{align}
which contains 4 non-trivial equations
\begin{gather}
    H_x H_y + H_y H_z + H_z H_x = 8\pi\rho \label{EFE1},\\
    \frac{\ddot{b}}{b}+ H_y H_z + \frac{\ddot{c}}{c} = -8\pi p \label{EFE2},\\
    \frac{\ddot{c}}{c}+ H_z H_x + \frac{\ddot{a}}{a} = -8\pi p \label{EFE3},\\
    \frac{\ddot{a}}{a}+ H_x H_y + \frac{\ddot{b}}{b} = -8\pi p \label{EFE4}.
\end{gather}
Rearranging \cref{EFE2,EFE3,EFE4} gives the following system of equations
\begin{gather}
    \frac{\ddot{a}}{a} = -4\pi p + \frac{1}{2}(H_y H_z) - \frac{1}{2}\left[H_x(H_y+H_z)\right] \label{EFE5},\\
    \frac{\ddot{b}}{b} = -4\pi p + \frac{1}{2}(H_z H_x) - \frac{1}{2}\left[H_y(H_z+H_x)\right] \label{EFE6},\\
    \frac{\ddot{c}}{c} = -4\pi p + \frac{1}{2}(H_x H_y) - \frac{1}{2}\left[H_z(H_x+H_y)\right] \label{EFE7}.
\end{gather}
Note that the present-day scale factors are, without loss of generality, defined to be unity ($a_0=b_0=c_0=1$), and the subscript `0' denotes the present values. Moreover, a flat universe is assumed such that \cref{EFE1} implies \cite{Russell_2014,Ng_2025}
\begin{gather}
    \begin{split}
        \rho_{r0} + \rho_{m0} + \rho_{\Lambda0} &= \rho_{\text{crit}0} \\
        &= \frac{H_{x0} H_{y0} + H_{y0} H_{z0} + H_{z0} H_{x0}}{8\pi}.
    \end{split}
\end{gather}

\subsection{Angular Acoustic Scale}
Following \cite{Ng_2025}, under the coordinate system in which \cref{Bianchi metric} is defined, sound waves at a polar angle $\Theta$ and an azimuthal angle $\phi$ propagate out to an ellipsoidal surface at photon decoupling, with semi-axes
\begin{gather}
    L_x = \int_0^{t_*} \frac{c_s(t)dt}{a(t)},\\
    L_y = \int_0^{t_*} \frac{c_s(t)dt}{b(t)},\\
    L_z = \int_0^{t_*} \frac{c_s(t)dt}{c(t)},\\
    c_s(t) = \frac{1}{\sqrt{3(1+R(t))}} = \frac{1}{\sqrt{3(1+\frac{3\rho_b(t)}{4\rho_r(t)})}}.
\end{gather}
Here, $\rho_b$ is the baryon density, and $t_*$ is the last scattering time of \ac{CMB} photons, i.e., the time when the optical depth equals unity. If the universe is isotropic, an observer would find this sound horizon projecting a circular ring with radius $r_*$ on his/her celestial sphere. Instead, if the universe is anisotropic, one would observe an elliptical ring with an enclosed area \cite{Vickers_1996}

\begin{gather}
    \text{Area} = \pi\sqrt{L_y^2 L_z^2 C_\phi^2 S_\Theta^2 + L_x^2 L_z^2 S_\phi^2 S_\Theta^2 + L_x^2 L_y^2 C_\Theta^2},
\intertext{where}
    \begin{aligned}
    C_\Theta &= \cos\Theta, &C_\phi &= \cos\phi,\\
    S_\Theta &= \sin\Theta, &\text{and } S_\phi &= \sin\phi.
    \end{aligned}
\end{gather}

An effective sound horizon can thus be defined
\begin{gather}
    r_* = \sqrt{\frac{\text{Area}}{\pi}}.
\end{gather}
On the other hand, by considering the null geodesics, the transverse comoving distance is given by \cite{Ng_2025, Koivisto_2008}
\begin{gather}
    D_M(\Theta, \phi) = \int_{t_*}^{t_0} \frac{dt}{\sqrt{a^2 C_\phi^2 S_\Theta^2 + b^2 S_\phi^2 S_\Theta^2 + c^2 C_\Theta^2}}.\label{D_M}
\end{gather}
Therefore, at the angular position ($\Theta,\phi$), the angular acoustic scale is
\begin{gather}
    \theta_*(\Theta,\phi)=\frac{r_*(\Theta,\phi)}{D_M(\Theta, \phi)}.
\end{gather}

We highlight several key features of the system. First, the last scattering time exhibits isotropy, which arises directly from the isotropic nature of both the baryon number density $n_b$ and the electron fraction $X_e = n_e/(n_p + n_\textrm{H})$. The isotropy of energy densities is evident from \cref{rho_a}, as anisotropies only influence energy densities through their `dilution' effects, while the densities themselves remain isotropic. However, the Area (and consequently $r_*$), along with $D_M$ and $\theta_*$, generally display anisotropy, with their directional variations being manifested through the trigonometric functions of $\Theta$ and $\phi$ that originate from the comoving distances.

\section{Methodology \label{Method}}
Ref. \cite{Yeung_2022} provides compelling evidence for large-scale anisotropy by analyzing the \ac{CMB} temperature power spectra. The authors partition the sky into 48 half-skies using the \texttt{HEALPix} \footnote{https://healpix.sourceforge.io/} scheme \cite{Gorski_2005} and perform \ac{MCMC} analyses on the half-sky power spectra to extract the cosmological parameters $\set{\Omega_{b} h^2, \Omega_{c} h^2, n_s, 100\theta_\mathrm{MC}, \tau, \ln{(10^{10}A_s)}}$ \cite{Yeung_2022}. Notably, they identify a distinct dipole-like variation in $100\theta_\mathrm{MC}$ \cite{Yeung_2022}, a parameter that is approximately equal to $100\theta_*$ and remains rather invariant across different cosmologies \cite{Planck_2018}, unlike other cosmological parameters. Hence, we study the feasibility of explaining the observed anisotropy in $\theta_\mathrm{MC}$ with the Bianchi Type I model based on Ref. \cite{Yeung_2022}'s approach.

We solve the evolution of the scale factors using the LSODA method \cite{Hindmarsh_1983, Petzold_1983}. The last scattering time of \ac{CMB} photons $t_*$ is calculated with a modified \texttt{CLASS} \cite{Blas_2011} as outlined in Ref. \cite{Ng_2025}. Instead of $\theta_\mathrm{MC}$, we first generate maps of $\theta_*(\Theta, \phi)$ with \texttt{healpy}, the Python package for \texttt{HEALPix} \cite{Zonca_2019, Gorski_2005}. We divide the sky into 192 pixels ($N_\mathrm{side}=4$ with RING ordering) and assign the value of $\theta_*$ for each pixel to be the pixel center value. This gives a rough idea of how $\theta_*$ varies with directions.

However, it is important to note that Ref. \cite{Yeung_2022} fitted the half-sky power spectra for the parameter $100\theta_\mathrm{MC}$. Consequently, the reported values of $100\theta_\mathrm{MC}$ represent some averages over half-skies which may not be directly comparable to the $\theta_*$ map we discussed above. Moreover, one must consider the possible rotation between the axes of the galactic coordinate system employed in Ref. \cite{Yeung_2022} and the anisotropic axes of the Bianchi Type I metric. Thus, we introduce three Euler angles $\psi_1, \psi_2, \psi_3$ with the `ZYX' convention to include such degrees of freedom. Since we do not have the \ac{CMB} power spectra calculated under the Bianchi Type I metric, we adopt the closest alternative by examining half-sky averages of $\theta_*$. Specifically, for a given pixel center $(\Theta_\mathrm{obs}, \phi_\mathrm{obs})$ from the observation in Ref. \cite{Yeung_2022}, we mask out the opposite hemisphere in the 192-pixel $\theta_*$ map and compute the half-sky average $\langle\theta_*\rangle_h$ as follows

\begin{gather}
    \langle\theta_*\rangle_h (\Theta_\mathrm{obs},\phi_\mathrm{obs})=\frac{1}{N_\text{unmasked}}\sum_{i\in\set{\text{unmasked}}}\theta_{*,i}\ ,
\end{gather}
where $i$ runs through the unmasked pixels in the 192-pixel map with pixel values of $\theta_{*,i}$, and $N_\text{unmasked}$ is the number of unmasked pixels which varies with $(\Theta_\mathrm{obs}, \phi_\mathrm{obs})$. Note that we are taking arithmetic means of $\theta_{*,i}$, as the \texttt{healpy} pixels have equal area \cite{Zonca_2019, Gorski_2005}. By repeating the process for the 48 pairs of ($\Theta_\mathrm{obs}, \phi_\mathrm{obs}$), we obtain a 48-pixel map of the $\langle\theta_*\rangle_h$ for comparison with Ref. \cite{Yeung_2022}. To verify that our pipeline does not produce artificial patterns beyond numerical accuracy, we apply our pipeline to the $\Lambda$ cold dark matter ($\Lambda$CDM) model. The results are shown in \cref{verify-pipeline}, where the sky maps are highly uniform, as expected.

% To facilitate comparison with the observed 48-pixel map, we down-sample the map to 48 pixels whenever comparison is needed. This approach ensures that our analysis is not constrained by resolution limitations, as we initially work with a map of significantly higher resolution than necessary.

In this work, we focus on the pattern in the directional variation of $\theta_*$, rather than their absolute values. Hence, we do not aim to find the best fit of the cosmological parameters and Euler angles. Instead, we will show that the Bianchi Type I model cannot reproduce the observed $100\theta_\mathrm{MC}$ directional dependency, regardless of the choice of parameters. Therefore, we consider five sets of cosmological parameters listed in \cref{Cosmo_param_table} alongside eight sets of Euler angles (\cref{Euler_ang_table}), with the asymmetry parameters $\alpha_y$ and $\alpha_z$ defined as
\begin{gather}
    \alpha_y = \frac{H_{y0}}{H_{x0}}-1 \quad \text{and} \quad \alpha_z = \frac{H_{z0}}{H_{x0}}-1.
\end{gather}
In particular, we consider asymmetry parameters of the order \num{E-10} which give the present-day \ac{CMB} temperature anisotropies $\lesssim \num{2E-5}$, consistent with observation \cite{Planck_2018}.

\begin{table}
    \centering
    \begin{tabularx}{0.8\columnwidth}
        {>{\raggedright\arraybackslash\hsize=0.7\hsize\linewidth=\hsize}X 
        >{\centering\arraybackslash\hsize=1.15\hsize\linewidth=\hsize}X
        >{\centering\arraybackslash\hsize=1.15\hsize\linewidth=\hsize}X}
        \toprule\toprule
        Set & $\alpha_y\ (\times \num{4E-10})$ & $\alpha_z\ (\times \num{4E-10})$ \\
        \midrule
        A1 & \num{1} & \num{0}\\
        A2 & \num{0} & \num{1}\\
        A3 & \num{1} & \num{1}\\
        A4 & \num{1} & \num{-1}\\
        A5 & \num{-1} & \num{-1}\\
        \bottomrule\bottomrule
    \end{tabularx}
    \caption{Values of asymmetry parameters used in this work. All parameter sets assume $(\omega_{b0}, \omega_{c0}, H_{x0}) = (0.0220, 0.1157, \SI{68.15}{\km\per\s\per\mega\pc})$ with $\omega_{b0}$ ($\omega_{c0}$) being the physical baryon density (cold dark matter) $8\pi\rho_{b0}(\rho_{c0})/3$. \label{Cosmo_param_table}}
\end{table}

\begin{table}
    \centering
    \begin{tabularx}{\columnwidth}
        {>{\raggedright\arraybackslash\hsize=1.1\hsize\linewidth=\hsize}X 
        >{\centering\arraybackslash\hsize=1.1\hsize\linewidth=\hsize}X
        >{\centering\arraybackslash\hsize=1.1\hsize\linewidth=\hsize}X
        >{\centering\arraybackslash\hsize=1.1\hsize\linewidth=\hsize}X|
        >{\centering\arraybackslash\hsize=0.2\hsize\linewidth=\hsize}X 
        >{\raggedright\arraybackslash\hsize=1.1\hsize\linewidth=\hsize}X 
        >{\centering\arraybackslash\hsize=1.1\hsize\linewidth=\hsize}X
        >{\centering\arraybackslash\hsize=1.1\hsize\linewidth=\hsize}X
        >{\centering\arraybackslash\hsize=1.1\hsize\linewidth=\hsize}X}
        \toprule\toprule
        Set & $\psi_1$ & $\psi_2$ & $\psi_3$ & & Set & $\psi_1$ & $\psi_2$ & $\psi_3$\\
        \midrule
        E1 & \num{0} & \num{0} & \num{0} & & E5 & $\dfrac{\pi}{3}$ & $\dfrac{\pi}{6}$ & \num{0} \\[3.5ex]
        E2 & \num{0} & \num{0} & $\dfrac{\pi}{3}$ & & E6 & $\dfrac{\pi}{3}$ & \num{0} & $\dfrac{5\pi}{3}$ \\[3.5ex]
        E3 & \num{0} & $\dfrac{\pi}{3}$ & \num{0} & & E7 & \num{0} & $\dfrac{\pi}{6}$ & $\dfrac{5\pi}{3}$ \\[3.5ex]
        E4 & $\dfrac{\pi}{3}$ & \num{0} & \num{0} & & E8 & $\dfrac{\pi}{3}$ & $\dfrac{\pi}{6}$ & $\dfrac{5\pi}{3}$ \\[3ex]
        \bottomrule\bottomrule
    \end{tabularx}
    \caption{Sets of Euler angles used with their respective labels. \label{Euler_ang_table}}
\end{table}

\section{Results and Discussion \label{Results}}
\begin{figure}
    \centering
    \includegraphics[width=1\linewidth]{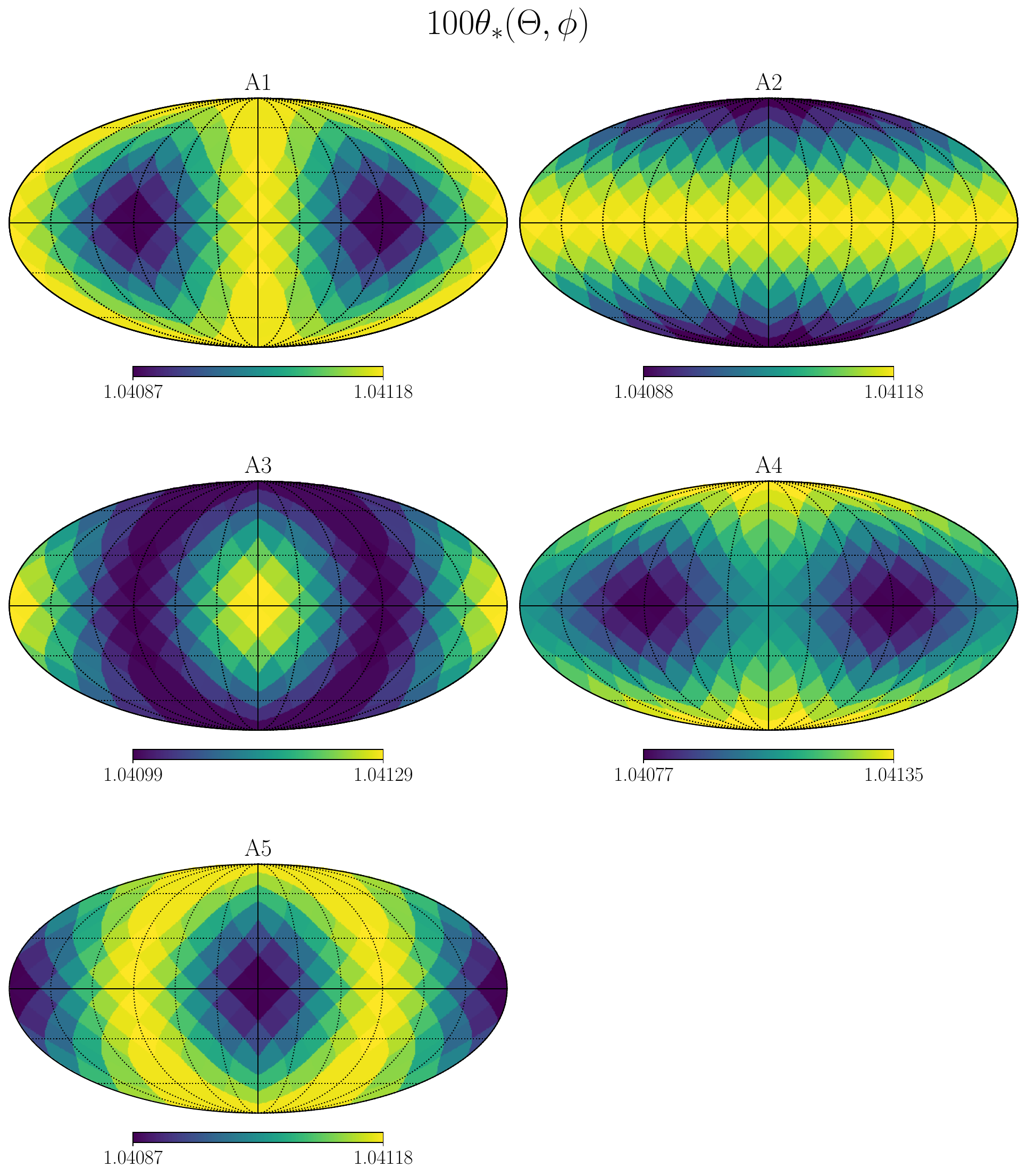}
    \caption{Values of $100\theta_*$ at different sky positions $(\Theta,\phi)$ generated with the five sets of asymmetry parameters, A1 to A5 (see \cref{Cosmo_param_table}). \label{theta_A1-5}}
\end{figure}

\cref{theta_A1-5} shows the directional variations of $100\theta_*$ under various sets of asymmetry parameters. For the sets A1 and A2, the distributions of $100\theta_*$ exhibit a clear axially symmetric pattern as expected. Notably, the $100\theta_*$ distributions consistently display parity symmetry across all asymmetry parameter sets, meaning that antipodal points share identical $100\theta_*$ values. This symmetry persists regardless of coordinate rotations or variations in the asymmetry parameters, which only alter the degree of anisotropy. Hence, the Bianchi Type I model generally produces a $\theta_*$ distribution that is even in parity. In fact, the parity symmetry can be observed from the parity-even metric and the modified evolution equations \cref{EFE5,EFE6,EFE7} which preserve it. This is in strong disagreement with the dipole distribution observed in Ref. \cite{Yeung_2022} which is odd in parity.

\begin{figure}
    \centering
    \includegraphics[width=1\linewidth]{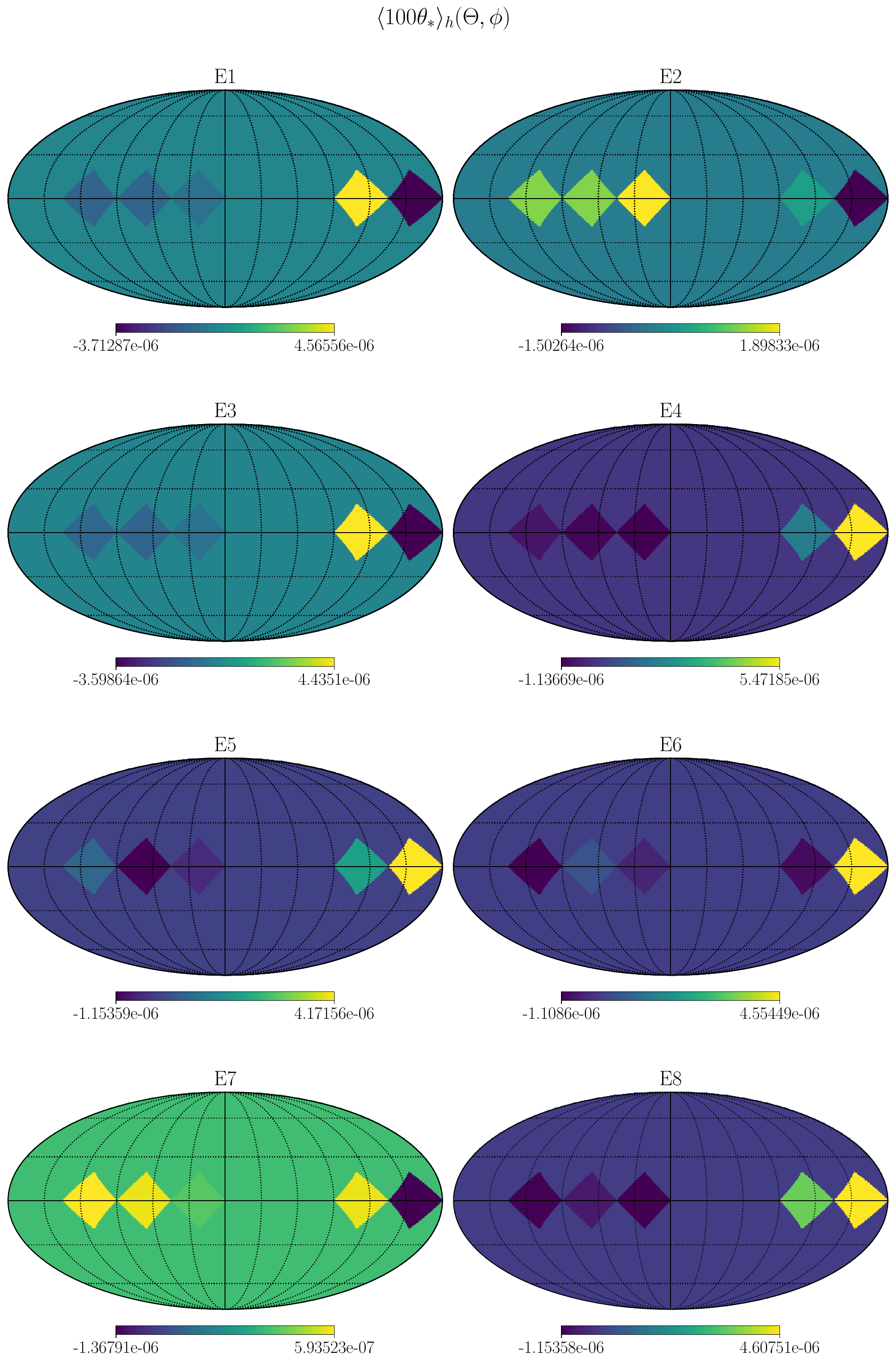}
    \caption{Variations of $\langle100\theta_*\rangle_h$ for half-skies centered at different sky positions $(\Theta,\phi)$, corresponding to the Euler angle sets E1 to E8 (see \cref{Euler_ang_table}), using the asymmetry parameter set A1. A constant of 1.04108 has been subtracted from all maps. \label{avg_theta_A1_E1-8}}
\end{figure}

\begin{figure}
    \centering
    \includegraphics[width=1\linewidth]{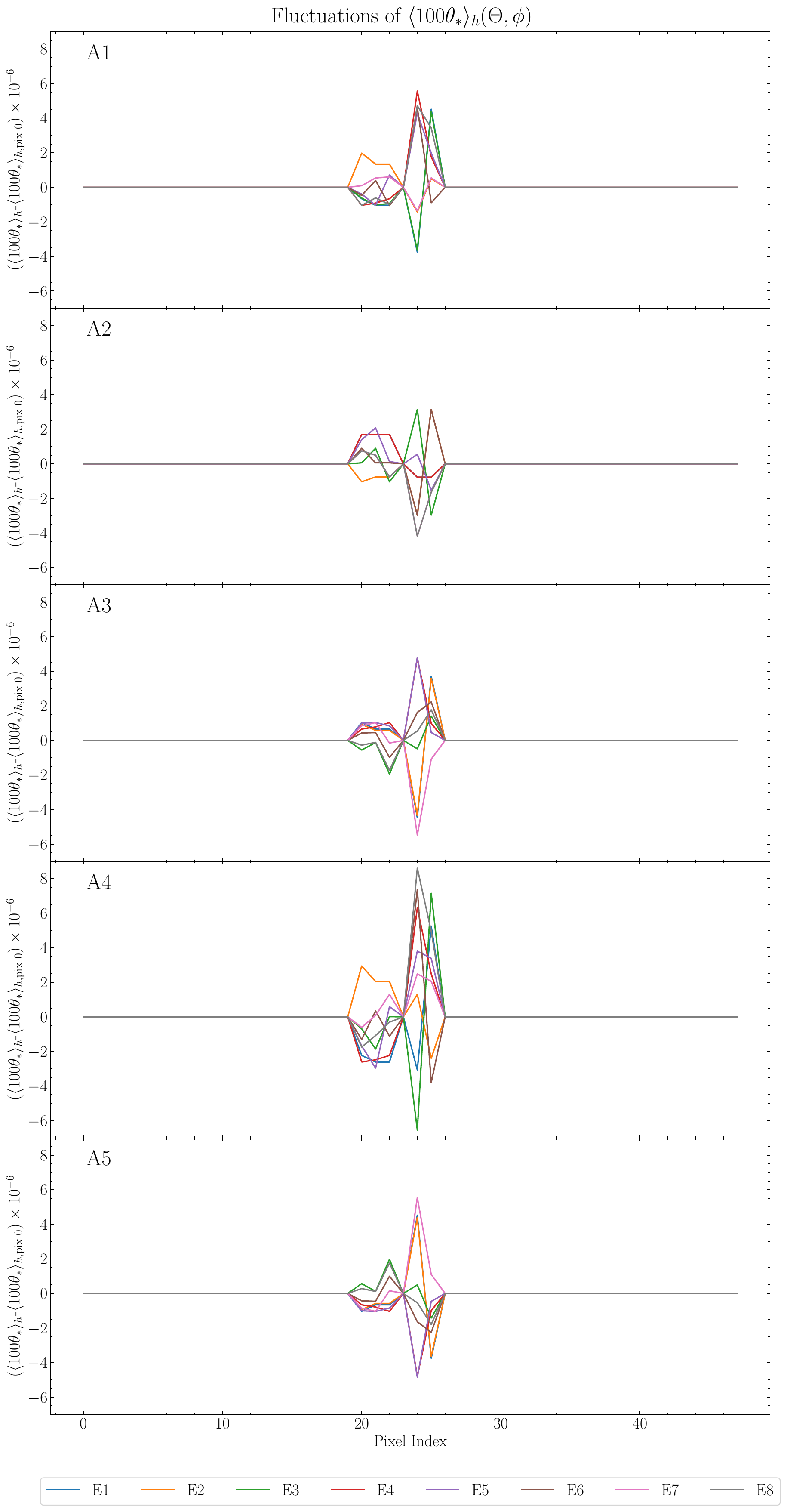}
    \caption{Fluctuations of $\langle100\theta_*\rangle_h$ plotted against the pixel index in RING ordering for various asymmetry parameter sets  (A1 to A5 from top to bottom, respectively). The colors represent the sets of Euler angles considered. \label{avg_theta_A1-8_E1-8}}
\end{figure}

As discussed in \cref{Method}, directly comparing the calculated $100\theta_*$ at each sky position with the observed value using \ac{MCMC} on half-sky spectra may not be appropriate. Instead, we study the half-sky average $\langle\theta_*\rangle_h$ to approximate the $\theta_*$ value derived from half-sky spectra. \cref{avg_theta_A1_E1-8} shows the directional dependency of $\langle100\theta_*\rangle_h$ in a universe characterized by the asymmetry parameter set A1 under various rotations. We observe that $\langle100\theta_*\rangle_h$ is highly uniform across all directions, with discrepancies on the equator remaining at $\lesssim\num{5e-6}$ level irrespective of the applied rotation. Such kind of defects stems from the masking of a finite-resolution sky map. Namely, the number of masked pixels fluctuates when the center of the mask is near the equator. Nonetheless, the uniformity exhibited is consistent across all sets of asymmetry parameters and Euler angles, as demonstrated in \cref{avg_theta_A1-8_E1-8}, which plots $\langle100\theta_*\rangle_h$ against the pixel index instead. The results confirm that, except for minor numerical artifacts, $\langle100\theta_*\rangle_h$ is homogeneous across all pixels which is in strong contrast to the dipole distribution reported \cite{Yeung_2022}. This holds true regardless of the asymmetry parameters and Euler angles. Therefore, the Bianchi Type I model cannot account for the dipole variation in $\theta_*$ observed in Ref. \cite{Yeung_2022}.

In recent years, given the increasing evidence of large-scale anisotropies in the cosmological observables \cite{Hansen_2004, Eriksen_2004, Bernui_2006, Bernui_2008, Bielewicz_2005, deOliveira-Costa_2004, Tegmark_2003, Wiaux_2006, Naselsky_2012, Zhao_2014, Cheng_2016, Aluri_2017, Bunn_2000, Eriksen_2004, Bernui_2007, Hansen_2009, Kashlinsky_2008, Watkins_2009, Hutsemekers_2005,Javanmardi_2015, Krishnan_2022, Luongo_2022, Hu_2023,Antoniou_2010, Zhao_2016, Perivolaropoulos_2022} as well as the challenges faced by the standard $\Lambda$CDM paradigm such as the Hubble tension \cite{Perivolaropoulos_2022, Planck_2018, Riess_2022}, many authors have turned their attention towards the Bianchi Type I metric \cite{Akarsu_2019, Shekh_2020, Sarmah_2022, Yadav_2023, Koussour_2023, Koivisto_2008, Hertzberg_Loeb_2024, Akarsu_2010, Adhav_2012, Sharif_2011, Shekh_2020, Saha_2001, Orjuela_2024, Ng_2025}. Our results show that the Bianchi Type I model cannot account for dipole-like large-scale anisotropies such as that in the observed $100\theta_\mathrm{MC}$ \cite{Yeung_2022}.

Here, we list some possible smoking gun signals of the Bianchi Type I model. The most promising feature would be the detection of an energetically anisotropic but parity-even \ac{CvB} \cite{Hertzberg_Loeb_2024,Ng_2025}. The associated anisotropic strength should be even stronger than that of \ac{CMB} since the decoupling time of neutrino is earlier than the photon decoupling time and the fact that the universe isotropizes for the Bianchi Type I model as time evolves \cite{Wald_1983, Akarsu_2019,Hertzberg_Loeb_2024}. A promising experiment aiming to detect the \ac{CvB} directly is the PTOLEMY experiment \cite{Rossi_2024}. Although the detection of \ac{CvB} may be imminent, the current experiment cannot detect the neutrino incident direction and thus lacks the capability to detect the anisotropy of the \ac{CvB}. Refs. \cite{Ciscar_2024, Herrera_2025}
propose an alternative method to detect the \ac{CvB} by considering cosmic ray-boosted \ac{CvB} which could be detected and traced by the IceCube experiment. However, the boosting process may hide the inherent anisotropy of the \ac{CvB}. A more practical observation of large-scale anisotropy is the localized measurement of the \ac{CMB} or \ac{BAO} angular acoustic angle. In \cref{theta_A1-5}, we already show that the Bianchi Type I metric produces a distinct parity-even pattern in $\theta_*$ and, by extension, the \ac{BAO} (inverse) angular acoustic scales. One may perform localized measurements of the \ac{CMB} or \ac{BAO} angular acoustic scales by trading off lower multipole information, which is irrelevant for our purpose, and compare them with the prediction of the Bianchi Type I cosmology. One can impose severe constraints on the anisotropy parameters in the Bianchi Type I model if the anisotropic pattern disagrees with the predictions in \cref{theta_A1-5}. Suppose localized measurements of $100\theta_*$ find an anisotropic pattern inconsistent with \cref{theta_A1-5} which has an amplitude of \num{e-4}. The anisotropy arising from the Bianchi Type I model must then be smaller than \num{e-4}, implying the asymmetry parameters $\lesssim\num{5e-10}$. Moreover, if more parity-even anomalies are discovered, they would provide more compelling evidence for the Bianchi Type I model. On the other hand, local effects may be responsible for the observed anisotropies.

If the Bianchi Type I model is indeed ruled out, one may consider more complex anisotropic models instead. For instance, Refs. \cite{Krishnan_2023,Allahyari_2025} have proposed a dipole cosmology model, which is built upon the Bianchi Type V or VII$_h$ metric. The authors suggest that such a model allows for cosmic flow, which can produce a dipole-like distribution in observables through a tilted energy-momentum tensor \cite{Krishnan_2023,Allahyari_2025}. On the other hand, many authors \cite{Ashtekar_2009, Wilson_2010, Akarsu_2010_B3, Rodrigues_2012, Smith_2025} are considering other types of anisotropic models, including those from the Bianchi family, while incorporating novel physics such as alternative gravity theories. However, it is unclear whether any of these models could produce the anisotropic $\theta_\mathrm{MC}$ observed in Ref. \cite{Yeung_2022}. Lastly, one cannot rule out the possibility that the observed anisotropies are due to some unknown systematic errors. For instance, \cite{Hu_2020, Boubel_2024} re-analyzed the available data and suggested that there is no statistically significant anisotropy associated with Supernovae and quasars. More effort is needed to forge a consensus on the significance of anisotropy in the observed data. 

\section{Conclusion\label{Conclusion}}
In this paper, we extend the work of Ref. \cite{Ng_2025} to a fully asymmetric Bianchi Type I universe and calculate the corresponding $100\theta_*$ and its half-sky average, $\langle100\theta_*\rangle_h$. We compare our theoretical prediction with the measured $100\theta_\mathrm{MC}$ anisotropy \cite{Yeung_2022}. By considering several sets of cosmological parameters and Euler angles, we 
\begin{enumerate}[(i)]
    \item show that values of $100\theta_*$ exhibit parity-even symmetry which disagrees with the observed parity-odd dipole distribution of $100\theta_\mathrm{MC}$ values
    \item illustrate that the half-sky averaged $\langle100\theta_*\rangle_h$ predicted by the Bianchi Type I model is highly uniform across different directions, for asymmetry parameters of order \num{e-10}, a level constrained by the observed CMB temperature anisotropies, so it cannot account for the observed dipole distribution of $100\theta_\mathrm{MC}$.
\end{enumerate}

\begin{acknowledgments}
    We acknowledge The Chinese University of Hong Kong Central Research Computing Cluster for their computational resources. This research is supported by grants from the Research Grants Council of the Hong Kong Special Administrative Region, China, under project Nos. AoE/P-404/18 and 14300223. The Python packages \texttt{NumPy} \cite{Numpy_2020}, \texttt{SciPy} \cite{Scipy_2020}, \texttt{Numba} \cite{Numba_2015}, \texttt{healpy} \cite{Zonca_2019,Gorski_2005}, and \texttt{Matplotlib} \cite{Matplotlib_2007} are used in this work.
\end{acknowledgments}

\appendix
\section{Verifying the Pipeline \label[appendix]{verify-pipeline}}
\begin{figure}
    \centering
    \includegraphics[width=1\linewidth]{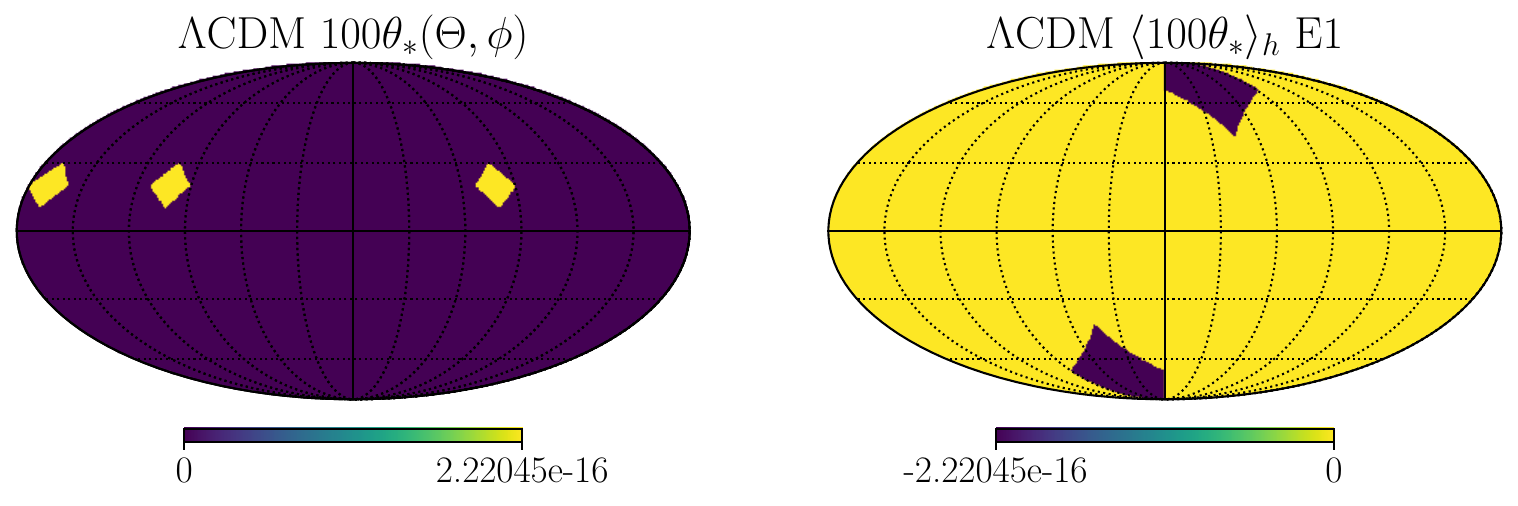}
    \caption{Sanity check of our pipeline with the $\Lambda$CDM model, showing variations of $100\theta_*$ (left panel) and $\langle100\theta_*\rangle_h$ (right panel). Note that both panels are subtracted by a constant of 1.0411.}
    \label{LCDM_plot}
\end{figure}

We verify that our pipeline does not introduce artificial patterns by testing it with the $\Lambda$CDM model. In the left panel of \cref{LCDM_plot}, we show the directional variation of $100\theta_*$ under the $\Lambda$CDM model. Note the high uniformity down to numerical precision, as expected under the model that assumes \ac{CP}, in strong contrast with \cref{theta_A1-5}. We plot the variation of $\langle 100\theta_*\rangle_h$ on the right panel of \cref{LCDM_plot}. The sky map remains highly uniform due to the $\Lambda$CDM model's assumption of isotropy. Additionally, numerical artifacts near the equators in \cref{avg_theta_A1_E1-8} are absent despite fluctuations in the number of masked pixels. This is due to the high uniformity of $100\theta_*$, which makes differences between different averaging of  pixels negligible. Overall, the pipeline is robust to numerical precision, except for tiny artifacts near the equators in \cref{avg_theta_A1_E1-8}.

\bibliography{References}

%apsrev4-2.bst 2019-01-14 (MD) hand-edited version of apsrev4-1.bst
%Control: key (0)
%Control: author (8) initials jnrlst
%Control: editor formatted (1) identically to author
%Control: production of article title (0) allowed
%Control: page (0) single
%Control: year (1) truncated
%Control: production of eprint (0) enabled
\begin{thebibliography}{66}%
\makeatletter
\providecommand \@ifxundefined [1]{%
 \@ifx{#1\undefined}
}%
\providecommand \@ifnum [1]{%
 \ifnum #1\expandafter \@firstoftwo
 \else \expandafter \@secondoftwo
 \fi
}%
\providecommand \@ifx [1]{%
 \ifx #1\expandafter \@firstoftwo
 \else \expandafter \@secondoftwo
 \fi
}%
\providecommand \natexlab [1]{#1}%
\providecommand \enquote  [1]{``#1''}%
\providecommand \bibnamefont  [1]{#1}%
\providecommand \bibfnamefont [1]{#1}%
\providecommand \citenamefont [1]{#1}%
\providecommand \href@noop [0]{\@secondoftwo}%
\providecommand \href [0]{\begingroup \@sanitize@url \@href}%
\providecommand \@href[1]{\@@startlink{#1}\@@href}%
\providecommand \@@href[1]{\endgroup#1\@@endlink}%
\providecommand \@sanitize@url [0]{\catcode `\\12\catcode `\$12\catcode `\&12\catcode `\#12\catcode `\^12\catcode `\_12\catcode `\%12\relax}%
\providecommand \@@startlink[1]{}%
\providecommand \@@endlink[0]{}%
\providecommand \url  [0]{\begingroup\@sanitize@url \@url }%
\providecommand \@url [1]{\endgroup\@href {#1}{\urlprefix }}%
\providecommand \urlprefix  [0]{URL }%
\providecommand \Eprint [0]{\href }%
\providecommand \doibase [0]{https://doi.org/}%
\providecommand \selectlanguage [0]{\@gobble}%
\providecommand \bibinfo  [0]{\@secondoftwo}%
\providecommand \bibfield  [0]{\@secondoftwo}%
\providecommand \translation [1]{[#1]}%
\providecommand \BibitemOpen [0]{}%
\providecommand \bibitemStop [0]{}%
\providecommand \bibitemNoStop [0]{.\EOS\space}%
\providecommand \EOS [0]{\spacefactor3000\relax}%
\providecommand \BibitemShut  [1]{\csname bibitem#1\endcsname}%
\let\auto@bib@innerbib\@empty
%</preamble>
\bibitem [{\citenamefont {Hansen}\ \emph {et~al.}(2004)\citenamefont {Hansen}, \citenamefont {Banday},\ and\ \citenamefont {Górski}}]{Hansen_2004}%
  \BibitemOpen
  \bibfield  {author} {\bibinfo {author} {\bibfnamefont {F.~K.}\ \bibnamefont {Hansen}}, \bibinfo {author} {\bibfnamefont {A.~J.}\ \bibnamefont {Banday}},\ and\ \bibinfo {author} {\bibfnamefont {K.~M.}\ \bibnamefont {Górski}},\ }\bibfield  {title} {\bibinfo {title} {{Testing the cosmological principle of isotropy: local power-spectrum estimates of the {WMAP} data}},\ }\href {https://doi.org/10.1111/j.1365-2966.2004.08229.x} {\bibfield  {journal} {\bibinfo  {journal} {Mon. Not. R. Astron. Soc.}\ }\textbf {\bibinfo {volume} {354}},\ \bibinfo {pages} {641} (\bibinfo {year} {2004})}\BibitemShut {NoStop}%
\bibitem [{\citenamefont {Eriksen}\ \emph {et~al.}(2004)\citenamefont {Eriksen}, \citenamefont {Hansen}, \citenamefont {Banday}, \citenamefont {Górski},\ and\ \citenamefont {Lilje}}]{Eriksen_2004}%
  \BibitemOpen
  \bibfield  {author} {\bibinfo {author} {\bibfnamefont {H.~K.}\ \bibnamefont {Eriksen}}, \bibinfo {author} {\bibfnamefont {F.~K.}\ \bibnamefont {Hansen}}, \bibinfo {author} {\bibfnamefont {A.~J.}\ \bibnamefont {Banday}}, \bibinfo {author} {\bibfnamefont {K.~M.}\ \bibnamefont {Górski}},\ and\ \bibinfo {author} {\bibfnamefont {P.~B.}\ \bibnamefont {Lilje}},\ }\bibfield  {title} {\bibinfo {title} {Asymmetries in the cosmic microwave background anisotropy field},\ }\href {https://doi.org/10.1086/382267} {\bibfield  {journal} {\bibinfo  {journal} {Astrophys. J.}\ }\textbf {\bibinfo {volume} {605}},\ \bibinfo {pages} {14} (\bibinfo {year} {2004})}\BibitemShut {NoStop}%
\bibitem [{\citenamefont {{Bernui, A.}}\ \emph {et~al.}(2006)\citenamefont {{Bernui, A.}}, \citenamefont {{Villela, T.}}, \citenamefont {{Wuensche, C. A.}}, \citenamefont {{Leonardi, R.}},\ and\ \citenamefont {{Ferreira, I.}}}]{Bernui_2006}%
  \BibitemOpen
  \bibfield  {author} {\bibinfo {author} {\bibnamefont {{Bernui, A.}}}, \bibinfo {author} {\bibnamefont {{Villela, T.}}}, \bibinfo {author} {\bibnamefont {{Wuensche, C. A.}}}, \bibinfo {author} {\bibnamefont {{Leonardi, R.}}},\ and\ \bibinfo {author} {\bibnamefont {{Ferreira, I.}}},\ }\bibfield  {title} {\bibinfo {title} {On the cosmic microwave background large-scale angular correlations},\ }\href {https://doi.org/10.1051/0004-6361:20054243} {\bibfield  {journal} {\bibinfo  {journal} {Astron. Astrophys.}\ }\textbf {\bibinfo {volume} {454}},\ \bibinfo {pages} {409} (\bibinfo {year} {2006})}\BibitemShut {NoStop}%
\bibitem [{\citenamefont {Bernui}(2008)}]{Bernui_2008}%
  \BibitemOpen
  \bibfield  {author} {\bibinfo {author} {\bibfnamefont {A.}~\bibnamefont {Bernui}},\ }\bibfield  {title} {\bibinfo {title} {Anomalous {CMB} north-south asymmetry},\ }\href {https://doi.org/10.1103/PhysRevD.78.063531} {\bibfield  {journal} {\bibinfo  {journal} {Phys. Rev. D}\ }\textbf {\bibinfo {volume} {78}},\ \bibinfo {pages} {063531} (\bibinfo {year} {2008})}\BibitemShut {NoStop}%
\bibitem [{\citenamefont {Bielewicz}\ \emph {et~al.}(2005)\citenamefont {Bielewicz}, \citenamefont {Eriksen}, \citenamefont {Banday}, \citenamefont {Górski},\ and\ \citenamefont {Lilje}}]{Bielewicz_2005}%
  \BibitemOpen
  \bibfield  {author} {\bibinfo {author} {\bibfnamefont {P.}~\bibnamefont {Bielewicz}}, \bibinfo {author} {\bibfnamefont {H.~K.}\ \bibnamefont {Eriksen}}, \bibinfo {author} {\bibfnamefont {A.~J.}\ \bibnamefont {Banday}}, \bibinfo {author} {\bibfnamefont {K.~M.}\ \bibnamefont {Górski}},\ and\ \bibinfo {author} {\bibfnamefont {P.~B.}\ \bibnamefont {Lilje}},\ }\bibfield  {title} {\bibinfo {title} {Multipole vector anomalies in the first-year {WMAP} data: A cut-sky analysis},\ }\href {https://doi.org/10.1086/497263} {\bibfield  {journal} {\bibinfo  {journal} {Astrophys. J.}\ }\textbf {\bibinfo {volume} {635}},\ \bibinfo {pages} {750} (\bibinfo {year} {2005})}\BibitemShut {NoStop}%
\bibitem [{\citenamefont {de~Oliveira-Costa}\ \emph {et~al.}(2004)\citenamefont {de~Oliveira-Costa}, \citenamefont {Tegmark}, \citenamefont {Zaldarriaga},\ and\ \citenamefont {Hamilton}}]{deOliveira-Costa_2004}%
  \BibitemOpen
  \bibfield  {author} {\bibinfo {author} {\bibfnamefont {A.}~\bibnamefont {de~Oliveira-Costa}}, \bibinfo {author} {\bibfnamefont {M.}~\bibnamefont {Tegmark}}, \bibinfo {author} {\bibfnamefont {M.}~\bibnamefont {Zaldarriaga}},\ and\ \bibinfo {author} {\bibfnamefont {A.}~\bibnamefont {Hamilton}},\ }\bibfield  {title} {\bibinfo {title} {Significance of the largest scale {CMB} fluctuations in {WMAP}},\ }\href {https://doi.org/10.1103/PhysRevD.69.063516} {\bibfield  {journal} {\bibinfo  {journal} {Phys. Rev. D}\ }\textbf {\bibinfo {volume} {69}},\ \bibinfo {pages} {063516} (\bibinfo {year} {2004})}\BibitemShut {NoStop}%
\bibitem [{\citenamefont {Tegmark}\ \emph {et~al.}(2003)\citenamefont {Tegmark}, \citenamefont {de~Oliveira-Costa},\ and\ \citenamefont {Hamilton}}]{Tegmark_2003}%
  \BibitemOpen
  \bibfield  {author} {\bibinfo {author} {\bibfnamefont {M.}~\bibnamefont {Tegmark}}, \bibinfo {author} {\bibfnamefont {A.}~\bibnamefont {de~Oliveira-Costa}},\ and\ \bibinfo {author} {\bibfnamefont {A.~J.~S.}\ \bibnamefont {Hamilton}},\ }\bibfield  {title} {\bibinfo {title} {High resolution foreground cleaned {CMB} map from {WMAP}},\ }\href {https://doi.org/10.1103/PhysRevD.68.123523} {\bibfield  {journal} {\bibinfo  {journal} {Phys. Rev. D}\ }\textbf {\bibinfo {volume} {68}},\ \bibinfo {pages} {123523} (\bibinfo {year} {2003})}\BibitemShut {NoStop}%
\bibitem [{\citenamefont {Wiaux}\ \emph {et~al.}(2006)\citenamefont {Wiaux}, \citenamefont {Vielva}, \citenamefont {Mart\'{\i}nez-Gonz\'alez},\ and\ \citenamefont {Vandergheynst}}]{Wiaux_2006}%
  \BibitemOpen
  \bibfield  {author} {\bibinfo {author} {\bibfnamefont {Y.}~\bibnamefont {Wiaux}}, \bibinfo {author} {\bibfnamefont {P.}~\bibnamefont {Vielva}}, \bibinfo {author} {\bibfnamefont {E.}~\bibnamefont {Mart\'{\i}nez-Gonz\'alez}},\ and\ \bibinfo {author} {\bibfnamefont {P.}~\bibnamefont {Vandergheynst}},\ }\bibfield  {title} {\bibinfo {title} {Global universe anisotropy probed by the alignment of structures in the cosmic microwave background},\ }\href {https://doi.org/10.1103/PhysRevLett.96.151303} {\bibfield  {journal} {\bibinfo  {journal} {Phys. Rev. Lett.}\ }\textbf {\bibinfo {volume} {96}},\ \bibinfo {pages} {151303} (\bibinfo {year} {2006})}\BibitemShut {NoStop}%
\bibitem [{\citenamefont {Naselsky}\ \emph {et~al.}(2012)\citenamefont {Naselsky}, \citenamefont {Zhao}, \citenamefont {Kim},\ and\ \citenamefont {Chen}}]{Naselsky_2012}%
  \BibitemOpen
  \bibfield  {author} {\bibinfo {author} {\bibfnamefont {P.}~\bibnamefont {Naselsky}}, \bibinfo {author} {\bibfnamefont {W.}~\bibnamefont {Zhao}}, \bibinfo {author} {\bibfnamefont {J.}~\bibnamefont {Kim}},\ and\ \bibinfo {author} {\bibfnamefont {S.}~\bibnamefont {Chen}},\ }\bibfield  {title} {\bibinfo {title} {Is the cosmic microwave background asymmetry due to the kinematic dipole?},\ }\href {https://doi.org/10.1088/0004-637X/749/1/31} {\bibfield  {journal} {\bibinfo  {journal} {Astrophys. J.}\ }\textbf {\bibinfo {volume} {749}},\ \bibinfo {pages} {31} (\bibinfo {year} {2012})}\BibitemShut {NoStop}%
\bibitem [{\citenamefont {Zhao}(2014)}]{Zhao_2014}%
  \BibitemOpen
  \bibfield  {author} {\bibinfo {author} {\bibfnamefont {W.}~\bibnamefont {Zhao}},\ }\bibfield  {title} {\bibinfo {title} {Directional dependence of {CMB} parity asymmetry},\ }\href {https://doi.org/10.1103/PhysRevD.89.023010} {\bibfield  {journal} {\bibinfo  {journal} {Phys. Rev. D}\ }\textbf {\bibinfo {volume} {89}},\ \bibinfo {pages} {023010} (\bibinfo {year} {2014})}\BibitemShut {NoStop}%
\bibitem [{\citenamefont {Cheng}\ \emph {et~al.}(2016)\citenamefont {Cheng}, \citenamefont {Zhao}, \citenamefont {Huang},\ and\ \citenamefont {Santos}}]{Cheng_2016}%
  \BibitemOpen
  \bibfield  {author} {\bibinfo {author} {\bibfnamefont {C.}~\bibnamefont {Cheng}}, \bibinfo {author} {\bibfnamefont {W.}~\bibnamefont {Zhao}}, \bibinfo {author} {\bibfnamefont {Q.-G.}\ \bibnamefont {Huang}},\ and\ \bibinfo {author} {\bibfnamefont {L.}~\bibnamefont {Santos}},\ }\bibfield  {title} {\bibinfo {title} {Preferred axis of {CMB} parity asymmetry in the masked maps},\ }\href {https://doi.org/10.1016/j.physletb.2016.04.030} {\bibfield  {journal} {\bibinfo  {journal} {Phys. Lett. B}\ }\textbf {\bibinfo {volume} {757}},\ \bibinfo {pages} {445} (\bibinfo {year} {2016})}\BibitemShut {NoStop}%
\bibitem [{\citenamefont {Aluri}\ \emph {et~al.}(2017)\citenamefont {Aluri}, \citenamefont {Ralston},\ and\ \citenamefont {Weltman}}]{Aluri_2017}%
  \BibitemOpen
  \bibfield  {author} {\bibinfo {author} {\bibfnamefont {P.~K.}\ \bibnamefont {Aluri}}, \bibinfo {author} {\bibfnamefont {J.~P.}\ \bibnamefont {Ralston}},\ and\ \bibinfo {author} {\bibfnamefont {A.}~\bibnamefont {Weltman}},\ }\bibfield  {title} {\bibinfo {title} {Alignments of parity even/odd-only multipoles in {CMB}},\ }\href {https://doi.org/10.1093/mnras/stx2112} {\bibfield  {journal} {\bibinfo  {journal} {Mon. Not. R. Astron. Soc.}\ }\textbf {\bibinfo {volume} {472}},\ \bibinfo {pages} {2410} (\bibinfo {year} {2017})}\BibitemShut {NoStop}%
\bibitem [{\citenamefont {Bunn}\ and\ \citenamefont {Scott}(2000)}]{Bunn_2000}%
  \BibitemOpen
  \bibfield  {author} {\bibinfo {author} {\bibfnamefont {E.~F.}\ \bibnamefont {Bunn}}\ and\ \bibinfo {author} {\bibfnamefont {D.}~\bibnamefont {Scott}},\ }\bibfield  {title} {\bibinfo {title} {A preferred-direction statistic for sky maps},\ }\href {https://doi.org/10.1051/0004-6361:20054243} {\bibfield  {journal} {\bibinfo  {journal} {Mon. Not. R. Astron. Soc.}\ }\textbf {\bibinfo {volume} {313}},\ \bibinfo {pages} {331} (\bibinfo {year} {2000})}\BibitemShut {NoStop}%
\bibitem [{\citenamefont {Bernui}\ \emph {et~al.}(2007)\citenamefont {Bernui}, \citenamefont {Mota}, \citenamefont {Reboucas},\ and\ \citenamefont {Tavakol}}]{Bernui_2007}%
  \BibitemOpen
  \bibfield  {author} {\bibinfo {author} {\bibfnamefont {A.}~\bibnamefont {Bernui}}, \bibinfo {author} {\bibfnamefont {B.}~\bibnamefont {Mota}}, \bibinfo {author} {\bibfnamefont {M.~J.}\ \bibnamefont {Reboucas}},\ and\ \bibinfo {author} {\bibfnamefont {R.}~\bibnamefont {Tavakol}},\ }\bibfield  {title} {\bibinfo {title} {Mapping the large-scale anisotropy in the {WMAP} data},\ }\href {https://doi.org/10.1051/0004-6361:20065585} {\bibfield  {journal} {\bibinfo  {journal} {Astron. Astrophys.}\ }\textbf {\bibinfo {volume} {464}},\ \bibinfo {pages} {479} (\bibinfo {year} {2007})}\BibitemShut {NoStop}%
\bibitem [{\citenamefont {Hansen}\ \emph {et~al.}(2009)\citenamefont {Hansen}, \citenamefont {Banday}, \citenamefont {Gorski}, \citenamefont {Eriksen},\ and\ \citenamefont {Lilje}}]{Hansen_2009}%
  \BibitemOpen
  \bibfield  {author} {\bibinfo {author} {\bibfnamefont {F.}~\bibnamefont {Hansen}}, \bibinfo {author} {\bibfnamefont {A.}~\bibnamefont {Banday}}, \bibinfo {author} {\bibfnamefont {K.}~\bibnamefont {Gorski}}, \bibinfo {author} {\bibfnamefont {H.}~\bibnamefont {Eriksen}},\ and\ \bibinfo {author} {\bibfnamefont {P.}~\bibnamefont {Lilje}},\ }\bibfield  {title} {\bibinfo {title} {Power asymmetry in cosmic microwave background fluctuations from full sky to sub-degree scales: is the universe isotropic?},\ }\href {https://doi.org/0004-637X/704/2/1448} {\bibfield  {journal} {\bibinfo  {journal} {Astrophys. J.}\ }\textbf {\bibinfo {volume} {704}},\ \bibinfo {pages} {1448} (\bibinfo {year} {2009})}\BibitemShut {NoStop}%
\bibitem [{\citenamefont {Kashlinsky}\ \emph {et~al.}(2008)\citenamefont {Kashlinsky}, \citenamefont {Atrio-Barandela}, \citenamefont {Kocevski},\ and\ \citenamefont {Ebeling}}]{Kashlinsky_2008}%
  \BibitemOpen
  \bibfield  {author} {\bibinfo {author} {\bibfnamefont {A.}~\bibnamefont {Kashlinsky}}, \bibinfo {author} {\bibfnamefont {F.}~\bibnamefont {Atrio-Barandela}}, \bibinfo {author} {\bibfnamefont {D.}~\bibnamefont {Kocevski}},\ and\ \bibinfo {author} {\bibfnamefont {H.}~\bibnamefont {Ebeling}},\ }\bibfield  {title} {\bibinfo {title} {A measurement of large-scale peculiar velocities of clusters of galaxies: results and cosmological implications},\ }\href {https://doi.org/10.1086/592947} {\bibfield  {journal} {\bibinfo  {journal} {Astrophys. J.}\ }\textbf {\bibinfo {volume} {686}},\ \bibinfo {pages} {L49} (\bibinfo {year} {2008})}\BibitemShut {NoStop}%
\bibitem [{\citenamefont {Watkins}\ \emph {et~al.}(2009)\citenamefont {Watkins}, \citenamefont {Feldman},\ and\ \citenamefont {Hudson}}]{Watkins_2009}%
  \BibitemOpen
  \bibfield  {author} {\bibinfo {author} {\bibfnamefont {R.}~\bibnamefont {Watkins}}, \bibinfo {author} {\bibfnamefont {H.~A.}\ \bibnamefont {Feldman}},\ and\ \bibinfo {author} {\bibfnamefont {M.~J.}\ \bibnamefont {Hudson}},\ }\bibfield  {title} {\bibinfo {title} {Consistently large cosmic flows on scales of 100 h- 1 mpc: a challenge for the standard {$\Lambda$CDM} cosmology},\ }\href {https://doi.org/10.1111/j.1365-2966.2008.14089.x} {\bibfield  {journal} {\bibinfo  {journal} {Mon. Not. R. Astron. Soc.}\ }\textbf {\bibinfo {volume} {392}},\ \bibinfo {pages} {743} (\bibinfo {year} {2009})}\BibitemShut {NoStop}%
\bibitem [{\citenamefont {Hutsem{\'e}kers}\ \emph {et~al.}(2005)\citenamefont {Hutsem{\'e}kers}, \citenamefont {Cabanac}, \citenamefont {Lamy},\ and\ \citenamefont {Sluse}}]{Hutsemekers_2005}%
  \BibitemOpen
  \bibfield  {author} {\bibinfo {author} {\bibfnamefont {D.}~\bibnamefont {Hutsem{\'e}kers}}, \bibinfo {author} {\bibfnamefont {R.}~\bibnamefont {Cabanac}}, \bibinfo {author} {\bibfnamefont {H.}~\bibnamefont {Lamy}},\ and\ \bibinfo {author} {\bibfnamefont {D.}~\bibnamefont {Sluse}},\ }\bibfield  {title} {\bibinfo {title} {Mapping extreme-scale alignments of quasar polarization vectors},\ }\href {https://doi.org/10.1051/0004-6361%3A20053337} {\bibfield  {journal} {\bibinfo  {journal} {Astron. Astrophys.}\ }\textbf {\bibinfo {volume} {441}},\ \bibinfo {pages} {915} (\bibinfo {year} {2005})}\BibitemShut {NoStop}%
\bibitem [{\citenamefont {Javanmardi}\ \emph {et~al.}(2015)\citenamefont {Javanmardi}, \citenamefont {Porciani}, \citenamefont {Kroupa},\ and\ \citenamefont {Pflamm-Altenburg}}]{Javanmardi_2015}%
  \BibitemOpen
  \bibfield  {author} {\bibinfo {author} {\bibfnamefont {B.}~\bibnamefont {Javanmardi}}, \bibinfo {author} {\bibfnamefont {C.}~\bibnamefont {Porciani}}, \bibinfo {author} {\bibfnamefont {P.}~\bibnamefont {Kroupa}},\ and\ \bibinfo {author} {\bibfnamefont {J.}~\bibnamefont {Pflamm-Altenburg}},\ }\bibfield  {title} {\bibinfo {title} {Probing the isotropy of cosmic acceleration traced by {Type Ia} supernovae},\ }\href {https://doi.org/10.1088/0004-637X/810/1/47} {\bibfield  {journal} {\bibinfo  {journal} {Astrophys. J.}\ }\textbf {\bibinfo {volume} {810}},\ \bibinfo {pages} {47} (\bibinfo {year} {2015})}\BibitemShut {NoStop}%
\bibitem [{\citenamefont {Krishnan}\ \emph {et~al.}(2022)\citenamefont {Krishnan}, \citenamefont {Mohayaee}, \citenamefont {Colg{\'a}in}, \citenamefont {Sheikh-Jabbari},\ and\ \citenamefont {Yin}}]{Krishnan_2022}%
  \BibitemOpen
  \bibfield  {author} {\bibinfo {author} {\bibfnamefont {C.}~\bibnamefont {Krishnan}}, \bibinfo {author} {\bibfnamefont {R.}~\bibnamefont {Mohayaee}}, \bibinfo {author} {\bibfnamefont {E.~{\'O}.}\ \bibnamefont {Colg{\'a}in}}, \bibinfo {author} {\bibfnamefont {M.}~\bibnamefont {Sheikh-Jabbari}},\ and\ \bibinfo {author} {\bibfnamefont {L.}~\bibnamefont {Yin}},\ }\bibfield  {title} {\bibinfo {title} {Hints of {FLRW} breakdown from supernovae},\ }\href {https://doi.org/10.1103/PhysRevD.105.063514} {\bibfield  {journal} {\bibinfo  {journal} {Phys. Rev. D}\ }\textbf {\bibinfo {volume} {105}},\ \bibinfo {pages} {063514} (\bibinfo {year} {2022})}\BibitemShut {NoStop}%
\bibitem [{\citenamefont {Luongo}\ \emph {et~al.}(2022)\citenamefont {Luongo}, \citenamefont {Muccino}, \citenamefont {Colg\'ain}, \citenamefont {Sheikh-Jabbari},\ and\ \citenamefont {Yin}}]{Luongo_2022}%
  \BibitemOpen
  \bibfield  {author} {\bibinfo {author} {\bibfnamefont {O.}~\bibnamefont {Luongo}}, \bibinfo {author} {\bibfnamefont {M.}~\bibnamefont {Muccino}}, \bibinfo {author} {\bibfnamefont {E.~O.}\ \bibnamefont {Colg\'ain}}, \bibinfo {author} {\bibfnamefont {M.~M.}\ \bibnamefont {Sheikh-Jabbari}},\ and\ \bibinfo {author} {\bibfnamefont {L.}~\bibnamefont {Yin}},\ }\bibfield  {title} {\bibinfo {title} {Larger {${H}_{0}$} values in the {CMB} dipole direction},\ }\href {https://doi.org/10.1103/PhysRevD.105.103510} {\bibfield  {journal} {\bibinfo  {journal} {Phys. Rev. D}\ }\textbf {\bibinfo {volume} {105}},\ \bibinfo {pages} {103510} (\bibinfo {year} {2022})}\BibitemShut {NoStop}%
\bibitem [{\citenamefont {Hu}\ \emph {et~al.}(2024)\citenamefont {Hu}, \citenamefont {Wang}, \citenamefont {Hu},\ and\ \citenamefont {Wang}}]{Hu_2023}%
  \BibitemOpen
  \bibfield  {author} {\bibinfo {author} {\bibfnamefont {J.}~\bibnamefont {Hu}}, \bibinfo {author} {\bibfnamefont {Y.}~\bibnamefont {Wang}}, \bibinfo {author} {\bibfnamefont {J.}~\bibnamefont {Hu}},\ and\ \bibinfo {author} {\bibfnamefont {F.}~\bibnamefont {Wang}},\ }\bibfield  {title} {\bibinfo {title} {Testing the cosmological principle with the pantheon+ sample and the region-fitting method},\ }\href {https://doi.org/10.1051/0004-6361/202347121} {\bibfield  {journal} {\bibinfo  {journal} {Astron. Astrophys.}\ }\textbf {\bibinfo {volume} {681}},\ \bibinfo {pages} {A88} (\bibinfo {year} {2024})}\BibitemShut {NoStop}%
\bibitem [{\citenamefont {Antoniou}\ and\ \citenamefont {Perivolaropoulos}(2010)}]{Antoniou_2010}%
  \BibitemOpen
  \bibfield  {author} {\bibinfo {author} {\bibfnamefont {I.}~\bibnamefont {Antoniou}}\ and\ \bibinfo {author} {\bibfnamefont {L.}~\bibnamefont {Perivolaropoulos}},\ }\bibfield  {title} {\bibinfo {title} {Searching for a cosmological preferred axis: {Union2} data analysis and comparison with other probes},\ }\href {https://doi.org/10.1088/1475-7516/2010/12/012} {\bibfield  {journal} {\bibinfo  {journal} {J. Cosmol. Astropart. Phys.}\ }\textbf {\bibinfo {volume} {2010}}\bibinfo  {number} { (12)},\ \bibinfo {pages} {012}}\BibitemShut {NoStop}%
\bibitem [{\citenamefont {Zhao}\ and\ \citenamefont {Santos}(2016)}]{Zhao_2016}%
  \BibitemOpen
\bibfield  {number} {  }\bibfield  {author} {\bibinfo {author} {\bibfnamefont {W.}~\bibnamefont {Zhao}}\ and\ \bibinfo {author} {\bibfnamefont {L.}~\bibnamefont {Santos}},\ }\bibfield  {title} {\bibinfo {title} {Preferred axis in cosmology},\ }\href@noop {} {\bibfield  {journal} {\bibinfo  {journal} {arXiv preprint arXiv:1604.05484}\ } (\bibinfo {year} {2016})},\ \Eprint {https://arxiv.org/abs/1604.05484} {arXiv:1604.05484 [astro-ph.CO]} \BibitemShut {NoStop}%
\bibitem [{\citenamefont {Perivolaropoulos}\ and\ \citenamefont {Skara}(2022)}]{Perivolaropoulos_2022}%
  \BibitemOpen
  \bibfield  {author} {\bibinfo {author} {\bibfnamefont {L.}~\bibnamefont {Perivolaropoulos}}\ and\ \bibinfo {author} {\bibfnamefont {F.}~\bibnamefont {Skara}},\ }\bibfield  {title} {\bibinfo {title} {Challenges for {$\Lambda$CDM}: An update},\ }\href {https://doi.org/10.1016/j.newar.2022.101659} {\bibfield  {journal} {\bibinfo  {journal} {New Astronomy Reviews}\ }\textbf {\bibinfo {volume} {95}},\ \bibinfo {pages} {101659} (\bibinfo {year} {2022})}\BibitemShut {NoStop}%
\bibitem [{\citenamefont {Axelsson}\ \emph {et~al.}(2013)\citenamefont {Axelsson}, \citenamefont {Fantaye}, \citenamefont {Hansen}, \citenamefont {Banday}, \citenamefont {Eriksen},\ and\ \citenamefont {Gorski}}]{Axelsson_2013}%
  \BibitemOpen
  \bibfield  {author} {\bibinfo {author} {\bibfnamefont {M.}~\bibnamefont {Axelsson}}, \bibinfo {author} {\bibfnamefont {Y.}~\bibnamefont {Fantaye}}, \bibinfo {author} {\bibfnamefont {F.}~\bibnamefont {Hansen}}, \bibinfo {author} {\bibfnamefont {A.}~\bibnamefont {Banday}}, \bibinfo {author} {\bibfnamefont {H.}~\bibnamefont {Eriksen}},\ and\ \bibinfo {author} {\bibfnamefont {K.}~\bibnamefont {Gorski}},\ }\bibfield  {title} {\bibinfo {title} {Directional dependence of {$\Lambda$CDM} cosmological parameters},\ }\href {https://doi.org/10.1088/2041-8205/773/1/L3} {\bibfield  {journal} {\bibinfo  {journal} {Astrophys. J. Lett.}\ }\textbf {\bibinfo {volume} {773}},\ \bibinfo {pages} {L3} (\bibinfo {year} {2013})}\BibitemShut {NoStop}%
\bibitem [{\citenamefont {Yeung}\ and\ \citenamefont {Chu}(2022)}]{Yeung_2022}%
  \BibitemOpen
  \bibfield  {author} {\bibinfo {author} {\bibfnamefont {S.}~\bibnamefont {Yeung}}\ and\ \bibinfo {author} {\bibfnamefont {M.-C.}\ \bibnamefont {Chu}},\ }\bibfield  {title} {\bibinfo {title} {Directional variations of cosmological parameters from the planck {CMB} data},\ }\href {https://doi.org/10.1103/PhysRevD.105.083508} {\bibfield  {journal} {\bibinfo  {journal} {Phys. Rev. D}\ }\textbf {\bibinfo {volume} {105}},\ \bibinfo {pages} {083508} (\bibinfo {year} {2022})}\BibitemShut {NoStop}%
\bibitem [{\citenamefont {Akarsu}\ \emph {et~al.}(2019)\citenamefont {Akarsu}, \citenamefont {Kumar}, \citenamefont {Sharma},\ and\ \citenamefont {Tedesco}}]{Akarsu_2019}%
  \BibitemOpen
  \bibfield  {author} {\bibinfo {author} {\bibfnamefont {{\"O}.}~\bibnamefont {Akarsu}}, \bibinfo {author} {\bibfnamefont {S.}~\bibnamefont {Kumar}}, \bibinfo {author} {\bibfnamefont {S.}~\bibnamefont {Sharma}},\ and\ \bibinfo {author} {\bibfnamefont {L.}~\bibnamefont {Tedesco}},\ }\bibfield  {title} {\bibinfo {title} {Constraints on a {Bianchi type I} spacetime extension of the standard {$\Lambda$ CDM} model},\ }\href {https://doi.org/10.1103/PhysRevD.100.023532} {\bibfield  {journal} {\bibinfo  {journal} {Phys. Rev. D}\ }\textbf {\bibinfo {volume} {100}},\ \bibinfo {pages} {023532} (\bibinfo {year} {2019})}\BibitemShut {NoStop}%
\bibitem [{\citenamefont {Shekh}\ and\ \citenamefont {Chirde}(2020)}]{Shekh_2020}%
  \BibitemOpen
  \bibfield  {author} {\bibinfo {author} {\bibfnamefont {S.~H.}\ \bibnamefont {Shekh}}\ and\ \bibinfo {author} {\bibfnamefont {V.~R.}\ \bibnamefont {Chirde}},\ }\bibfield  {title} {\bibinfo {title} {Accelerating bianchi type dark energy cosmological model with cosmic string in $f(t)$ gravity},\ }\bibfield  {journal} {\bibinfo  {journal} {Astrophysics and Space Science}\ }\textbf {\bibinfo {volume} {365}},\ \href {https://doi.org/10.1007/s10509-020-03772-y} {10.1007/s10509-020-03772-y} (\bibinfo {year} {2020})\BibitemShut {NoStop}%
\bibitem [{\citenamefont {Sarmah}\ and\ \citenamefont {Goswami}(2022)}]{Sarmah_2022}%
  \BibitemOpen
  \bibfield  {author} {\bibinfo {author} {\bibfnamefont {P.}~\bibnamefont {Sarmah}}\ and\ \bibinfo {author} {\bibfnamefont {U.~D.}\ \bibnamefont {Goswami}},\ }\bibfield  {title} {\bibinfo {title} {Bianchi type i model of universe with customized scale factors},\ }\bibfield  {journal} {\bibinfo  {journal} {Modern Physics Letters A}\ }\textbf {\bibinfo {volume} {37}},\ \href {https://doi.org/10.1142/s0217732322501346} {10.1142/s0217732322501346} (\bibinfo {year} {2022})\BibitemShut {NoStop}%
\bibitem [{\citenamefont {Koussour}\ \emph {et~al.}(2023)\citenamefont {Koussour}, \citenamefont {Shekh}, \citenamefont {Govender},\ and\ \citenamefont {Bennai}}]{Koussour_2023}%
  \BibitemOpen
  \bibfield  {author} {\bibinfo {author} {\bibfnamefont {M.}~\bibnamefont {Koussour}}, \bibinfo {author} {\bibfnamefont {S.}~\bibnamefont {Shekh}}, \bibinfo {author} {\bibfnamefont {M.}~\bibnamefont {Govender}},\ and\ \bibinfo {author} {\bibfnamefont {M.}~\bibnamefont {Bennai}},\ }\bibfield  {title} {\bibinfo {title} {Thermodynamical aspects of bianchi type-i universe in quadratic form of f(q) gravity and observational constraints},\ }\href {https://doi.org/10.1016/j.jheap.2022.11.002} {\bibfield  {journal} {\bibinfo  {journal} {Journal of High Energy Astrophysics}\ }\textbf {\bibinfo {volume} {37}},\ \bibinfo {pages} {15–24} (\bibinfo {year} {2023})}\BibitemShut {NoStop}%
\bibitem [{\citenamefont {Hertzberg}\ and\ \citenamefont {Loeb}(2024)}]{Hertzberg_Loeb_2024}%
  \BibitemOpen
  \bibfield  {author} {\bibinfo {author} {\bibfnamefont {M.~P.}\ \bibnamefont {Hertzberg}}\ and\ \bibinfo {author} {\bibfnamefont {A.}~\bibnamefont {Loeb}},\ }\bibfield  {title} {\bibinfo {title} {Constraints on an anisotropic universe},\ }\href {https://doi.org/10.1103/PhysRevD.109.083538} {\bibfield  {journal} {\bibinfo  {journal} {Phys. Rev. D}\ }\textbf {\bibinfo {volume} {109}},\ \bibinfo {pages} {083538} (\bibinfo {year} {2024})}\BibitemShut {NoStop}%
\bibitem [{\citenamefont {Ng}\ and\ \citenamefont {Chu}(2025)}]{Ng_2025}%
  \BibitemOpen
  \bibfield  {author} {\bibinfo {author} {\bibfnamefont {B.~H.-L.}\ \bibnamefont {Ng}}\ and\ \bibinfo {author} {\bibfnamefont {M.-C.}\ \bibnamefont {Chu}},\ }\bibfield  {title} {\bibinfo {title} {Constraining the locally rotationally symmetric bianchi type i model with self-consistent recombination history and observables},\ }\href {https://doi.org/10.1103/3njx-sy22} {\bibfield  {journal} {\bibinfo  {journal} {Physical Review D}\ }\textbf {\bibinfo {volume} {112}},\ \bibinfo {pages} {023553} (\bibinfo {year} {2025})}\BibitemShut {NoStop}%
\bibitem [{\citenamefont {Yadav}(2023)}]{Yadav_2023}%
  \BibitemOpen
  \bibfield  {author} {\bibinfo {author} {\bibfnamefont {V.}~\bibnamefont {Yadav}},\ }\bibfield  {title} {\bibinfo {title} {Measuring hubble constant in an anisotropic extension of {$\Lambda$CDM} model},\ }\href {https://doi.org/10.1016/j.dark.2023.101365} {\bibfield  {journal} {\bibinfo  {journal} {Physics of the Dark Universe}\ }\textbf {\bibinfo {volume} {42}},\ \bibinfo {pages} {101365} (\bibinfo {year} {2023})}\BibitemShut {NoStop}%
\bibitem [{\citenamefont {Koivisto}\ and\ \citenamefont {Mota}(2008)}]{Koivisto_2008}%
  \BibitemOpen
  \bibfield  {author} {\bibinfo {author} {\bibfnamefont {T.}~\bibnamefont {Koivisto}}\ and\ \bibinfo {author} {\bibfnamefont {D.~F.}\ \bibnamefont {Mota}},\ }\bibfield  {title} {\bibinfo {title} {Anisotropic dark energy: dynamics of the background and perturbations},\ }\href {https://doi.org/10.1088/1475-7516/2008/06/018} {\bibfield  {journal} {\bibinfo  {journal} {J. Cosmol. Astropart. Phys.}\ }\textbf {\bibinfo {volume} {2008}}\bibinfo  {number} { (06)},\ \bibinfo {pages} {018}}\BibitemShut {NoStop}%
\bibitem [{\citenamefont {Russell}\ \emph {et~al.}(2014)\citenamefont {Russell}, \citenamefont {Kılınç},\ and\ \citenamefont {Pashaev}}]{Russell_2014}%
  \BibitemOpen
\bibfield  {number} {  }\bibfield  {author} {\bibinfo {author} {\bibfnamefont {E.}~\bibnamefont {Russell}}, \bibinfo {author} {\bibfnamefont {C.~B.}\ \bibnamefont {Kılınç}},\ and\ \bibinfo {author} {\bibfnamefont {O.~K.}\ \bibnamefont {Pashaev}},\ }\bibfield  {title} {\bibinfo {title} {{Bianchi I model: an alternative way to model the present-day Universe}},\ }\href {https://doi.org/10.1093/mnras/stu932} {\bibfield  {journal} {\bibinfo  {journal} {Mon. Not. R. Astron. Soc.}\ }\textbf {\bibinfo {volume} {442}},\ \bibinfo {pages} {2331} (\bibinfo {year} {2014})}\BibitemShut {NoStop}%
\bibitem [{\citenamefont {Vickers}(1996)}]{Vickers_1996}%
  \BibitemOpen
  \bibfield  {author} {\bibinfo {author} {\bibfnamefont {G.}~\bibnamefont {Vickers}},\ }\bibfield  {title} {\bibinfo {title} {The projected areas of ellipsoids and cylinders},\ }\href {https://doi.org/10.1016/0032-5910(95)03049-2} {\bibfield  {journal} {\bibinfo  {journal} {Powder technology}\ }\textbf {\bibinfo {volume} {86}},\ \bibinfo {pages} {195} (\bibinfo {year} {1996})}\BibitemShut {NoStop}%
\bibitem [{\citenamefont {Gorski}\ \emph {et~al.}(2005)\citenamefont {Gorski}, \citenamefont {Hivon}, \citenamefont {Banday}, \citenamefont {Wandelt}, \citenamefont {Hansen}, \citenamefont {Reinecke},\ and\ \citenamefont {Bartelmann}}]{Gorski_2005}%
  \BibitemOpen
  \bibfield  {author} {\bibinfo {author} {\bibfnamefont {K.~M.}\ \bibnamefont {Gorski}}, \bibinfo {author} {\bibfnamefont {E.}~\bibnamefont {Hivon}}, \bibinfo {author} {\bibfnamefont {A.~J.}\ \bibnamefont {Banday}}, \bibinfo {author} {\bibfnamefont {B.~D.}\ \bibnamefont {Wandelt}}, \bibinfo {author} {\bibfnamefont {F.~K.}\ \bibnamefont {Hansen}}, \bibinfo {author} {\bibfnamefont {M.}~\bibnamefont {Reinecke}},\ and\ \bibinfo {author} {\bibfnamefont {M.}~\bibnamefont {Bartelmann}},\ }\bibfield  {title} {\bibinfo {title} {Healpix: A framework for high-resolution discretization and fast analysis of data distributed on the sphere},\ }\href {https://doi.org/10.1086/427976} {\bibfield  {journal} {\bibinfo  {journal} {Astrophys. J.}\ }\textbf {\bibinfo {volume} {622}},\ \bibinfo {pages} {759} (\bibinfo {year} {2005})}\BibitemShut {NoStop}%
\bibitem [{\citenamefont {{Aghanim, N.}}\ \emph {et~al.}(2020)\citenamefont {{Aghanim, N.}} \emph {et~al.}}]{Planck_2018}%
  \BibitemOpen
  \bibfield  {author} {\bibinfo {author} {\bibnamefont {{Aghanim, N.}}} \emph {et~al.} (\bibinfo {collaboration} {Planck}),\ }\bibfield  {title} {\bibinfo {title} {Planck 2018 results - vi. cosmological parameters},\ }\href {https://doi.org/10.1051/0004-6361/201833910} {\bibfield  {journal} {\bibinfo  {journal} {Astron. Astrophys.}\ }\textbf {\bibinfo {volume} {641}},\ \bibinfo {pages} {A6} (\bibinfo {year} {2020})}\BibitemShut {NoStop}%
\bibitem [{\citenamefont {Hindmarsh}(1983)}]{Hindmarsh_1983}%
  \BibitemOpen
  \bibfield  {author} {\bibinfo {author} {\bibfnamefont {A.~C.}\ \bibnamefont {Hindmarsh}},\ }\bibfield  {title} {\bibinfo {title} {Odepack, a systemized collection of ode solvers},\ }\href {https://cir.nii.ac.jp/crid/1572543025424393088} {\bibfield  {journal} {\bibinfo  {journal} {Scientific computing}\ } (\bibinfo {year} {1983})}\BibitemShut {NoStop}%
\bibitem [{\citenamefont {Petzold}(1983)}]{Petzold_1983}%
  \BibitemOpen
  \bibfield  {author} {\bibinfo {author} {\bibfnamefont {L.}~\bibnamefont {Petzold}},\ }\bibfield  {title} {\bibinfo {title} {Automatic selection of methods for solving stiff and nonstiff systems of ordinary differential equations},\ }\href {https://doi.org/10.1137/0904010} {\bibfield  {journal} {\bibinfo  {journal} {SIAM journal on scientific and statistical computing}\ }\textbf {\bibinfo {volume} {4}},\ \bibinfo {pages} {136} (\bibinfo {year} {1983})}\BibitemShut {NoStop}%
\bibitem [{\citenamefont {Blas}\ \emph {et~al.}(2011)\citenamefont {Blas}, \citenamefont {Lesgourgues},\ and\ \citenamefont {Tram}}]{Blas_2011}%
  \BibitemOpen
  \bibfield  {author} {\bibinfo {author} {\bibfnamefont {D.}~\bibnamefont {Blas}}, \bibinfo {author} {\bibfnamefont {J.}~\bibnamefont {Lesgourgues}},\ and\ \bibinfo {author} {\bibfnamefont {T.}~\bibnamefont {Tram}},\ }\bibfield  {title} {\bibinfo {title} {The cosmic linear anisotropy solving system ({CLASS}). part ii: approximation schemes},\ }\href {https://doi.org/10.1088/1475-7516/2011/07/034} {\bibfield  {journal} {\bibinfo  {journal} {J. Cosmol. Astropart. Phys.}\ }\textbf {\bibinfo {volume} {2011}}\bibinfo  {number} { (07)},\ \bibinfo {pages} {034}}\BibitemShut {NoStop}%
\bibitem [{\citenamefont {Zonca}\ \emph {et~al.}(2019)\citenamefont {Zonca}, \citenamefont {Singer}, \citenamefont {Lenz}, \citenamefont {Reinecke}, \citenamefont {Rosset}, \citenamefont {Hivon},\ and\ \citenamefont {Gorski}}]{Zonca_2019}%
  \BibitemOpen
\bibfield  {number} {  }\bibfield  {author} {\bibinfo {author} {\bibfnamefont {A.}~\bibnamefont {Zonca}}, \bibinfo {author} {\bibfnamefont {L.}~\bibnamefont {Singer}}, \bibinfo {author} {\bibfnamefont {D.}~\bibnamefont {Lenz}}, \bibinfo {author} {\bibfnamefont {M.}~\bibnamefont {Reinecke}}, \bibinfo {author} {\bibfnamefont {C.}~\bibnamefont {Rosset}}, \bibinfo {author} {\bibfnamefont {E.}~\bibnamefont {Hivon}},\ and\ \bibinfo {author} {\bibfnamefont {K.}~\bibnamefont {Gorski}},\ }\bibfield  {title} {\bibinfo {title} {healpy: equal area pixelization and spherical harmonics transforms for data on the sphere in python},\ }\href {https://doi.org/10.21105/joss.01298} {\bibfield  {journal} {\bibinfo  {journal} {Journal of Open Source Software}\ }\textbf {\bibinfo {volume} {4}},\ \bibinfo {pages} {1298} (\bibinfo {year} {2019})}\BibitemShut {NoStop}%
\bibitem [{\citenamefont {Riess}\ \emph {et~al.}(2022)\citenamefont {Riess}, \citenamefont {Yuan}, \citenamefont {Macri}, \citenamefont {Scolnic}, \citenamefont {Brout}, \citenamefont {Casertano}, \citenamefont {Jones}, \citenamefont {Murakami}, \citenamefont {Anand}, \citenamefont {Breuval} \emph {et~al.}}]{Riess_2022}%
  \BibitemOpen
  \bibfield  {author} {\bibinfo {author} {\bibfnamefont {A.~G.}\ \bibnamefont {Riess}}, \bibinfo {author} {\bibfnamefont {W.}~\bibnamefont {Yuan}}, \bibinfo {author} {\bibfnamefont {L.~M.}\ \bibnamefont {Macri}}, \bibinfo {author} {\bibfnamefont {D.}~\bibnamefont {Scolnic}}, \bibinfo {author} {\bibfnamefont {D.}~\bibnamefont {Brout}}, \bibinfo {author} {\bibfnamefont {S.}~\bibnamefont {Casertano}}, \bibinfo {author} {\bibfnamefont {D.~O.}\ \bibnamefont {Jones}}, \bibinfo {author} {\bibfnamefont {Y.}~\bibnamefont {Murakami}}, \bibinfo {author} {\bibfnamefont {G.~S.}\ \bibnamefont {Anand}}, \bibinfo {author} {\bibfnamefont {L.}~\bibnamefont {Breuval}}, \emph {et~al.},\ }\bibfield  {title} {\bibinfo {title} {A comprehensive measurement of the local value of the hubble constant with 1 km s- 1 mpc- 1 uncertainty from the hubble space telescope and the sh0es team},\ }\href {https://doi.org/10.3847/2041-8213/ac5c5b} {\bibfield  {journal} {\bibinfo  {journal} {Astrophys. J. Lett.}\ }\textbf {\bibinfo {volume}
  {934}},\ \bibinfo {pages} {L7} (\bibinfo {year} {2022})}\BibitemShut {NoStop}%
\bibitem [{\citenamefont {Akarsu}\ and\ \citenamefont {K{\i}l{\i}n{\c{c}}}(2010{\natexlab{a}})}]{Akarsu_2010}%
  \BibitemOpen
  \bibfield  {author} {\bibinfo {author} {\bibfnamefont {{\"O}.}~\bibnamefont {Akarsu}}\ and\ \bibinfo {author} {\bibfnamefont {C.~B.}\ \bibnamefont {K{\i}l{\i}n{\c{c}}}},\ }\bibfield  {title} {\bibinfo {title} {Lrs bianchi type i models with anisotropic dark energy and constant deceleration parameter},\ }\href {https://doi.org/10.1007/s10714-009-0821-y} {\bibfield  {journal} {\bibinfo  {journal} {General Relativity and Gravitation}\ }\textbf {\bibinfo {volume} {42}},\ \bibinfo {pages} {119} (\bibinfo {year} {2010}{\natexlab{a}})}\BibitemShut {NoStop}%
\bibitem [{\citenamefont {Adhav}(2012)}]{Adhav_2012}%
  \BibitemOpen
  \bibfield  {author} {\bibinfo {author} {\bibfnamefont {K.}~\bibnamefont {Adhav}},\ }\bibfield  {title} {\bibinfo {title} {Lrs bianchi type-i cosmological model in f (r, t) theory of gravity},\ }\href {https://doi.org/10.1007/s10509-011-0963-8} {\bibfield  {journal} {\bibinfo  {journal} {Astrophysics and space science}\ }\textbf {\bibinfo {volume} {339}},\ \bibinfo {pages} {365} (\bibinfo {year} {2012})}\BibitemShut {NoStop}%
\bibitem [{\citenamefont {Sharif}\ and\ \citenamefont {Rani}(2011)}]{Sharif_2011}%
  \BibitemOpen
  \bibfield  {author} {\bibinfo {author} {\bibfnamefont {M.}~\bibnamefont {Sharif}}\ and\ \bibinfo {author} {\bibfnamefont {S.}~\bibnamefont {Rani}},\ }\bibfield  {title} {\bibinfo {title} {F (t) models within bianchi type-i universe},\ }\href {https://doi.org/10.1142/S0217732311036127} {\bibfield  {journal} {\bibinfo  {journal} {Modern Physics Letters A}\ }\textbf {\bibinfo {volume} {26}},\ \bibinfo {pages} {1657} (\bibinfo {year} {2011})}\BibitemShut {NoStop}%
\bibitem [{\citenamefont {Saha}(2001)}]{Saha_2001}%
  \BibitemOpen
  \bibfield  {author} {\bibinfo {author} {\bibfnamefont {B.}~\bibnamefont {Saha}},\ }\bibfield  {title} {\bibinfo {title} {Spinor field in a bianchi type-i universe: regular solutions},\ }\href {https://doi.org/10.1103/PhysRevD.64.123501} {\bibfield  {journal} {\bibinfo  {journal} {Phys. Rev. D}\ }\textbf {\bibinfo {volume} {64}},\ \bibinfo {pages} {123501} (\bibinfo {year} {2001})}\BibitemShut {NoStop}%
\bibitem [{\citenamefont {Orjuela-Quintana}\ \emph {et~al.}(2024)\citenamefont {Orjuela-Quintana}, \citenamefont {Palacios-C{\'o}rdoba},\ and\ \citenamefont {Valenzuela-Toledo}}]{Orjuela_2024}%
  \BibitemOpen
  \bibfield  {author} {\bibinfo {author} {\bibfnamefont {J.~B.}\ \bibnamefont {Orjuela-Quintana}}, \bibinfo {author} {\bibfnamefont {J.~L.}\ \bibnamefont {Palacios-C{\'o}rdoba}},\ and\ \bibinfo {author} {\bibfnamefont {C.~A.}\ \bibnamefont {Valenzuela-Toledo}},\ }\bibfield  {title} {\bibinfo {title} {Late-time anisotropy sourced by a 2-form field non-minimally coupled to cold dark matter},\ }\href {https://doi.org/10.1016/j.dark.2024.101575} {\bibfield  {journal} {\bibinfo  {journal} {Physics of the Dark Universe}\ }\textbf {\bibinfo {volume} {46}},\ \bibinfo {pages} {101575} (\bibinfo {year} {2024})}\BibitemShut {NoStop}%
\bibitem [{\citenamefont {Wald}(1983)}]{Wald_1983}%
  \BibitemOpen
  \bibfield  {author} {\bibinfo {author} {\bibfnamefont {R.~M.}\ \bibnamefont {Wald}},\ }\bibfield  {title} {\bibinfo {title} {Asymptotic behavior of homogeneous cosmological models in the presence of a positive cosmological constant},\ }\href {https://doi.org/10.1103/PhysRevD.28.2118} {\bibfield  {journal} {\bibinfo  {journal} {Phys. Rev. D}\ }\textbf {\bibinfo {volume} {28}},\ \bibinfo {pages} {2118} (\bibinfo {year} {1983})}\BibitemShut {NoStop}%
\bibitem [{\citenamefont {Rossi}\ \emph {et~al.}(2024)\citenamefont {Rossi} \emph {et~al.}}]{Rossi_2024}%
  \BibitemOpen
  \bibfield  {author} {\bibinfo {author} {\bibfnamefont {N.}~\bibnamefont {Rossi}} \emph {et~al.} (\bibinfo {collaboration} {PTOLEMY}),\ }\bibfield  {title} {\bibinfo {title} {{Cosmic Neutrino Background detection with PTOLEMY}},\ }\href {https://doi.org/10.22323/1.449.0103} {\bibfield  {journal} {\bibinfo  {journal} {Proceedings of Science}\ }\textbf {\bibinfo {volume} {EPS-HEP2023}},\ \bibinfo {pages} {103} (\bibinfo {year} {2024})}\BibitemShut {NoStop}%
\bibitem [{\citenamefont {C{\'\i}scar-Monsalvatje}\ \emph {et~al.}(2024)\citenamefont {C{\'\i}scar-Monsalvatje}, \citenamefont {Herrera},\ and\ \citenamefont {Shoemaker}}]{Ciscar_2024}%
  \BibitemOpen
  \bibfield  {author} {\bibinfo {author} {\bibfnamefont {M.}~\bibnamefont {C{\'\i}scar-Monsalvatje}}, \bibinfo {author} {\bibfnamefont {G.}~\bibnamefont {Herrera}},\ and\ \bibinfo {author} {\bibfnamefont {I.~M.}\ \bibnamefont {Shoemaker}},\ }\bibfield  {title} {\bibinfo {title} {Upper limits on the cosmic neutrino background from cosmic rays},\ }\href {https://doi.org/10.1103/PhysRevD.110.063036} {\bibfield  {journal} {\bibinfo  {journal} {Phys. Rev. D}\ }\textbf {\bibinfo {volume} {110}},\ \bibinfo {pages} {063036} (\bibinfo {year} {2024})}\BibitemShut {NoStop}%
\bibitem [{\citenamefont {Herrera}\ \emph {et~al.}(2025)\citenamefont {Herrera}, \citenamefont {Horiuchi},\ and\ \citenamefont {Qi}}]{Herrera_2025}%
  \BibitemOpen
  \bibfield  {author} {\bibinfo {author} {\bibfnamefont {G.}~\bibnamefont {Herrera}}, \bibinfo {author} {\bibfnamefont {S.}~\bibnamefont {Horiuchi}},\ and\ \bibinfo {author} {\bibfnamefont {X.}~\bibnamefont {Qi}},\ }\bibfield  {title} {\bibinfo {title} {Diffuse boosted cosmic neutrino background},\ }\href {https://doi.org/10.1103/PhysRevD.111.063016} {\bibfield  {journal} {\bibinfo  {journal} {Phys. Rev. D}\ }\textbf {\bibinfo {volume} {111}},\ \bibinfo {pages} {063016} (\bibinfo {year} {2025})}\BibitemShut {NoStop}%
\bibitem [{\citenamefont {Krishnan}\ \emph {et~al.}(2023)\citenamefont {Krishnan}, \citenamefont {Mondol},\ and\ \citenamefont {Sheikh-Jabbari}}]{Krishnan_2023}%
  \BibitemOpen
  \bibfield  {author} {\bibinfo {author} {\bibfnamefont {C.}~\bibnamefont {Krishnan}}, \bibinfo {author} {\bibfnamefont {R.}~\bibnamefont {Mondol}},\ and\ \bibinfo {author} {\bibfnamefont {M.}~\bibnamefont {Sheikh-Jabbari}},\ }\bibfield  {title} {\bibinfo {title} {Dipole cosmology: the copernican paradigm beyond flrw},\ }\href {https://doi.org/https://doi.org/10.1088/1475-7516/2023/07/020} {\bibfield  {journal} {\bibinfo  {journal} {jcap}\ }\textbf {\bibinfo {volume} {2023}},\ \bibinfo {pages} {020} (\bibinfo {year} {2023})}\BibitemShut {NoStop}%
\bibitem [{\citenamefont {Allahyari}\ \emph {et~al.}(2025)\citenamefont {Allahyari}, \citenamefont {Ebrahimian}, \citenamefont {Mondol},\ and\ \citenamefont {Sheikh-Jabbari}}]{Allahyari_2025}%
  \BibitemOpen
  \bibfield  {author} {\bibinfo {author} {\bibfnamefont {A.}~\bibnamefont {Allahyari}}, \bibinfo {author} {\bibfnamefont {E.}~\bibnamefont {Ebrahimian}}, \bibinfo {author} {\bibfnamefont {R.}~\bibnamefont {Mondol}},\ and\ \bibinfo {author} {\bibfnamefont {M.}~\bibnamefont {Sheikh-Jabbari}},\ }\bibfield  {title} {\bibinfo {title} {Big bang in dipole cosmology},\ }\href {https://doi.org/10.1140/epjc/s10052-025-13799-6} {\bibfield  {journal} {\bibinfo  {journal} {The European Physical Journal C}\ }\textbf {\bibinfo {volume} {85}},\ \bibinfo {pages} {119} (\bibinfo {year} {2025})}\BibitemShut {NoStop}%
\bibitem [{\citenamefont {Ashtekar}\ and\ \citenamefont {Wilson-Ewing}(2009)}]{Ashtekar_2009}%
  \BibitemOpen
  \bibfield  {author} {\bibinfo {author} {\bibfnamefont {A.}~\bibnamefont {Ashtekar}}\ and\ \bibinfo {author} {\bibfnamefont {E.}~\bibnamefont {Wilson-Ewing}},\ }\bibfield  {title} {\bibinfo {title} {Loop quantum cosmology of bianchi type ii models},\ }\href {https://doi.org/10.1103/PhysRevD.80.123532} {\bibfield  {journal} {\bibinfo  {journal} {Phys. Rev. D}\ }\textbf {\bibinfo {volume} {80}},\ \bibinfo {pages} {123532} (\bibinfo {year} {2009})}\BibitemShut {NoStop}%
\bibitem [{\citenamefont {Wilson-Ewing}(2010)}]{Wilson_2010}%
  \BibitemOpen
  \bibfield  {author} {\bibinfo {author} {\bibfnamefont {E.}~\bibnamefont {Wilson-Ewing}},\ }\bibfield  {title} {\bibinfo {title} {Loop quantum cosmology of bianchi type ix models},\ }\href {https://doi.org/10.1103/PhysRevD.82.043508} {\bibfield  {journal} {\bibinfo  {journal} {Phys. Rev. D}\ }\textbf {\bibinfo {volume} {82}},\ \bibinfo {pages} {043508} (\bibinfo {year} {2010})}\BibitemShut {NoStop}%
\bibitem [{\citenamefont {Akarsu}\ and\ \citenamefont {K{\i}l{\i}n{\c{c}}}(2010{\natexlab{b}})}]{Akarsu_2010_B3}%
  \BibitemOpen
  \bibfield  {author} {\bibinfo {author} {\bibfnamefont {{\"O}.}~\bibnamefont {Akarsu}}\ and\ \bibinfo {author} {\bibfnamefont {C.~B.}\ \bibnamefont {K{\i}l{\i}n{\c{c}}}},\ }\bibfield  {title} {\bibinfo {title} {Bianchi type iii models with anisotropic dark energy},\ }\href {https://doi.org/10.1007/s10714-009-0878-7} {\bibfield  {journal} {\bibinfo  {journal} {General Relativity and Gravitation}\ }\textbf {\bibinfo {volume} {42}},\ \bibinfo {pages} {763} (\bibinfo {year} {2010}{\natexlab{b}})}\BibitemShut {NoStop}%
\bibitem [{\citenamefont {Rodrigues}\ \emph {et~al.}(2012)\citenamefont {Rodrigues}, \citenamefont {Houndjo}, \citenamefont {Saez-Gomez},\ and\ \citenamefont {Rahaman}}]{Rodrigues_2012}%
  \BibitemOpen
  \bibfield  {author} {\bibinfo {author} {\bibfnamefont {M.}~\bibnamefont {Rodrigues}}, \bibinfo {author} {\bibfnamefont {M.}~\bibnamefont {Houndjo}}, \bibinfo {author} {\bibfnamefont {D.}~\bibnamefont {Saez-Gomez}},\ and\ \bibinfo {author} {\bibfnamefont {F.}~\bibnamefont {Rahaman}},\ }\bibfield  {title} {\bibinfo {title} {Anisotropic universe models in f (t) gravity},\ }\href {https://doi.org/10.1103/PhysRevD.86.104059} {\bibfield  {journal} {\bibinfo  {journal} {Phys. Rev. D}\ }\textbf {\bibinfo {volume} {86}},\ \bibinfo {pages} {104059} (\bibinfo {year} {2012})}\BibitemShut {NoStop}%
\bibitem [{\citenamefont {Smith}\ \emph {et~al.}(2025)\citenamefont {Smith}, \citenamefont {Copi},\ and\ \citenamefont {Starkman}}]{Smith_2025}%
  \BibitemOpen
  \bibfield  {author} {\bibinfo {author} {\bibfnamefont {A.~F.}\ \bibnamefont {Smith}}, \bibinfo {author} {\bibfnamefont {C.~J.}\ \bibnamefont {Copi}},\ and\ \bibinfo {author} {\bibfnamefont {G.~D.}\ \bibnamefont {Starkman}},\ }\bibfield  {title} {\bibinfo {title} {Cosmological constraints on anisotropic thurston geometries},\ }\href {https://doi.org/10.1088/1475-7516/2025/01/005} {\bibfield  {journal} {\bibinfo  {journal} {jcap}\ }\textbf {\bibinfo {volume} {2025}},\ \bibinfo {pages} {005} (\bibinfo {year} {2025})}\BibitemShut {NoStop}%
\bibitem [{\citenamefont {Hu}\ \emph {et~al.}(2020)\citenamefont {Hu}, \citenamefont {Wang},\ and\ \citenamefont {Wang}}]{Hu_2020}%
  \BibitemOpen
  \bibfield  {author} {\bibinfo {author} {\bibfnamefont {J.}~\bibnamefont {Hu}}, \bibinfo {author} {\bibfnamefont {Y.}~\bibnamefont {Wang}},\ and\ \bibinfo {author} {\bibfnamefont {F.}~\bibnamefont {Wang}},\ }\bibfield  {title} {\bibinfo {title} {Testing cosmic anisotropy with pantheon sample and quasars at high redshifts},\ }\href {https://doi.org/10.1051/0004-6361/202038541} {\bibfield  {journal} {\bibinfo  {journal} {Astron. Astrophys.}\ }\textbf {\bibinfo {volume} {643}},\ \bibinfo {pages} {A93} (\bibinfo {year} {2020})}\BibitemShut {NoStop}%
\bibitem [{\citenamefont {Boubel}\ \emph {et~al.}(2024)\citenamefont {Boubel}, \citenamefont {Colless}, \citenamefont {Said},\ and\ \citenamefont {Staveley-Smith}}]{Boubel_2024}%
  \BibitemOpen
  \bibfield  {author} {\bibinfo {author} {\bibfnamefont {P.}~\bibnamefont {Boubel}}, \bibinfo {author} {\bibfnamefont {M.}~\bibnamefont {Colless}}, \bibinfo {author} {\bibfnamefont {K.}~\bibnamefont {Said}},\ and\ \bibinfo {author} {\bibfnamefont {L.}~\bibnamefont {Staveley-Smith}},\ }\bibfield  {title} {\bibinfo {title} {Testing anisotropic hubble expansion},\ }\href@noop {} {\bibfield  {journal} {\bibinfo  {journal} {arXiv preprint arXiv:2412.14607}\ } (\bibinfo {year} {2024})},\ \Eprint {https://arxiv.org/abs/2412.14607} {arXiv:2412.14607 [astro-ph.CO]} \BibitemShut {NoStop}%
\bibitem [{\citenamefont {Harris}\ \emph {et~al.}(2020)\citenamefont {Harris} \emph {et~al.}}]{Numpy_2020}%
  \BibitemOpen
  \bibfield  {author} {\bibinfo {author} {\bibfnamefont {C.~R.}\ \bibnamefont {Harris}} \emph {et~al.},\ }\bibfield  {title} {\bibinfo {title} {Array programming with {NumPy}},\ }\href {https://doi.org/10.1038/s41586-020-2649-2} {\bibfield  {journal} {\bibinfo  {journal} {Nature}\ }\textbf {\bibinfo {volume} {585}},\ \bibinfo {pages} {357} (\bibinfo {year} {2020})}\BibitemShut {NoStop}%
\bibitem [{\citenamefont {Virtanen}\ \emph {et~al.}(2020)\citenamefont {Virtanen} \emph {et~al.}}]{Scipy_2020}%
  \BibitemOpen
  \bibfield  {author} {\bibinfo {author} {\bibfnamefont {P.}~\bibnamefont {Virtanen}} \emph {et~al.},\ }\bibfield  {title} {\bibinfo {title} {{{SciPy} 1.0: Fundamental Algorithms for Scientific Computing in Python}},\ }\href {https://doi.org/10.1038/s41592-019-0686-2} {\bibfield  {journal} {\bibinfo  {journal} {Nature Methods}\ }\textbf {\bibinfo {volume} {17}},\ \bibinfo {pages} {261} (\bibinfo {year} {2020})}\BibitemShut {NoStop}%
\bibitem [{\citenamefont {Lam}\ \emph {et~al.}(2015)\citenamefont {Lam}, \citenamefont {Pitrou},\ and\ \citenamefont {Seibert}}]{Numba_2015}%
  \BibitemOpen
  \bibfield  {author} {\bibinfo {author} {\bibfnamefont {S.~K.}\ \bibnamefont {Lam}}, \bibinfo {author} {\bibfnamefont {A.}~\bibnamefont {Pitrou}},\ and\ \bibinfo {author} {\bibfnamefont {S.}~\bibnamefont {Seibert}},\ }\bibfield  {title} {\bibinfo {title} {Numba: A llvm-based python jit compiler},\ }in\ \href {https://doi.org/10.1145/2833157.2833162} {\emph {\bibinfo {booktitle} {Proceedings of the Second Workshop on the LLVM Compiler Infrastructure in HPC}}}\ (\bibinfo {year} {2015})\ pp.\ \bibinfo {pages} {1--6}\BibitemShut {NoStop}%
\bibitem [{\citenamefont {Hunter}(2007)}]{Matplotlib_2007}%
  \BibitemOpen
  \bibfield  {author} {\bibinfo {author} {\bibfnamefont {J.~D.}\ \bibnamefont {Hunter}},\ }\bibfield  {title} {\bibinfo {title} {Matplotlib: A 2d graphics environment},\ }\href {https://doi.org/10.1109/MCSE.2007.55} {\bibfield  {journal} {\bibinfo  {journal} {Computing in Science \& Engineering}\ }\textbf {\bibinfo {volume} {9}},\ \bibinfo {pages} {90} (\bibinfo {year} {2007})}\BibitemShut {NoStop}%
\end{thebibliography}%
\end{document}